\documentclass[sigconf, screen, nonacm]{acmart}

\AtBeginDocument{%
  }

\usepackage{amsmath}
\usepackage{algorithmic}
\usepackage[caption=false,font=footnotesize]{subfig}
\usepackage{tabularx}
\usepackage{textcomp}
\usepackage{verbatim}
\usepackage{listings}
\usepackage{stfloats}
\usepackage{multirow}
\usepackage{makecell}
\usepackage{threeparttable}
\usepackage{pifont}
\definecolor{DarkGreen}{rgb}{0.0, 0.5, 0.0}
\newcommand{\cmark}{{\color{DarkGreen}\ding{51}}}
\newcommand{\xmark}{{\color{red}\ding{55}}}
\usepackage{booktabs}
\usepackage{array}

\begin{document}

\title{Ten-Four: An Open-Source Fused Dot Product Unit for Mixed-Precision GPGPU Tensor Cores}

\author{Nikhil Rout}
\affiliation{%
  \institution{University of California, Los Angeles}
  \city{Los Angeles}
  \state{CA}
  \country{USA}
}
\email{nikhilrout97@ucla.edu}

\author{Blaise Tine}
\affiliation{%
  \institution{University of California, Los Angeles}
  \city{Los Angeles}
  \state{CA}
  \country{USA}
}
\email{blaisetine@cs.ucla.edu}

\begin{abstract}
Efficient mixed-precision MMA operations are critical for accelerating deep learning workloads on GPGPUs. However, existing open-source GPGPU Tensor Cores rely on discrete arithmetic units causing high latency, accumulated rounding errors, and poor resource utilization. To address these challenges, we propose Ten-Four, a configurable mixed-precision fused dot product (FEDP) unit microarchitecture implemented as part of the open-source RISC-V Vortex GPGPU's Tensor Core Unit extension. It supports low-precision multiplication in TF32, FP16, BF16, FP8, BF8, INT8 and INT4 with FP32/INT32 accumulation, native OCP Microscaling (MX), and sparse lane clock-gating for dynamic power reduction, while matching NVIDIA Tensor Core numerical accuracy. Ten-Four achieves 4-cycle latency at 300~MHz on the AMD Alveo U55C FPGA, delivering 2.7$\times$--7.9$\times$ speedup over Berkeley HardFloat and FPnew at less than 60\% the area cost. We further demonstrate its format extensibility by prototyping LNSF4, an NVFP4-variant matching its perplexity with 6.7\% fewer LUTs and 11.1\% lower dynamic power.
\end{abstract}



\maketitle

\section{Introduction}

As deep learning workloads continue to grow in scale and prevalence, General Matrix Multiply (GEMM) acceleration has become a critical priority for GPU designers due to its dominant role in model execution. For instance, profiling Llama 3.1 8B on NVIDIA B200 GPUs showed that over 80\% of the runtime is spent executing some variant of GEMM~\cite{taha2025blackwell}.
To mitigate this bottleneck, GPU vendors introduced dedicated throughput-focused matrix engines, such as NVIDIA Tensor Cores~\cite{nvidia2017volta} and AMD Matrix Cores~\cite{amd2021cdna2}, to execute specialized Warp-Matrix-Multiply-Accumulate (WMMA) and Matrix-Fused-Multiply-Add (MFMA) instructions, respectively.

Microbenchmarking these instructions has revealed key microarchitectural details of the underlying hardware~\cite{jia2018dissectingnvidiavoltagpu, 8695642, jia2019dissectingnvidiaturingt4, 9926299, 9931992, 10579250, luo2025dissectingnvidiahopperarchitecture, jarmusch2026microbenchmarkingnvidiasblackwellarchitecture}. Tensor Cores receive input sub-matrices \texttt{A}, \texttt{B}, and \texttt{C} directly from the SIMT Sub-Core register file, perform an \texttt{MxNxK} MMA, and write the resulting sub-matrix \texttt{D} back to the register file, much like the integer ALU, FPU, and LSU SIMD lanes they are placed alongside. To maximize throughput while maintaining numerical stability, the multiplicands \texttt{A} and \texttt{B} use lower precision formats (e.g. FP16/INT8), while the addend \texttt{C} and result \texttt{D} are in higher precision (FP32/INT32). These MMA operations substantially minimize power consumption and memory traffic by avoiding intermediate result storage. This enables greater compute density within the same die area and improves the evermore critical throughput/mm$^{2}$ and throughput/watt metrics in data-center GPUs.

While Tensor Cores in commercial GPUs have advanced remarkably in recent years (NVIDIA FP16 Tensor Core FLOPS increased eight-fold from Volta to Hopper generations~\cite{nvidia2022hopper}), the open-source GPGPU design space has lagged behind with sub-optimal prototypes delivering nominal throughput. This gap has only widened with the introduction of 2:4 structured sparsity since the Ampere generation~\cite{nvidia2020ampere, mishra2021acceleratingsparsedeepneural} and native hardware-accelerated OCP Microscaling (MX) format~\cite{ocp2023mx, rouhani2023microscalingdataformatsdeep} support in NVIDIA Blackwell~\cite{nvidia2024blackwell} and AMD CDNA 4~\cite{amd2025cdna4} architectures. This is largely a consequence of open-source designs' reliance on discrete floating-point arithmetic unit libraries, which introduce high latency, accumulated rounding errors, and poor resource utilization. For instance, both the Ventus GPGPU Tensor Core~\cite{10818098,  VentusTVLSI2025} and Virgo cluster-level systolic array matrix unit~\cite{10.1145/3676641.3716281} were built using discrete arithmetic modules from the Berkeley HardFloat~\cite{hauser2019hardfloat} library.
\begin{figure}[t]
    \centering
    \includegraphics[width=0.93\linewidth]{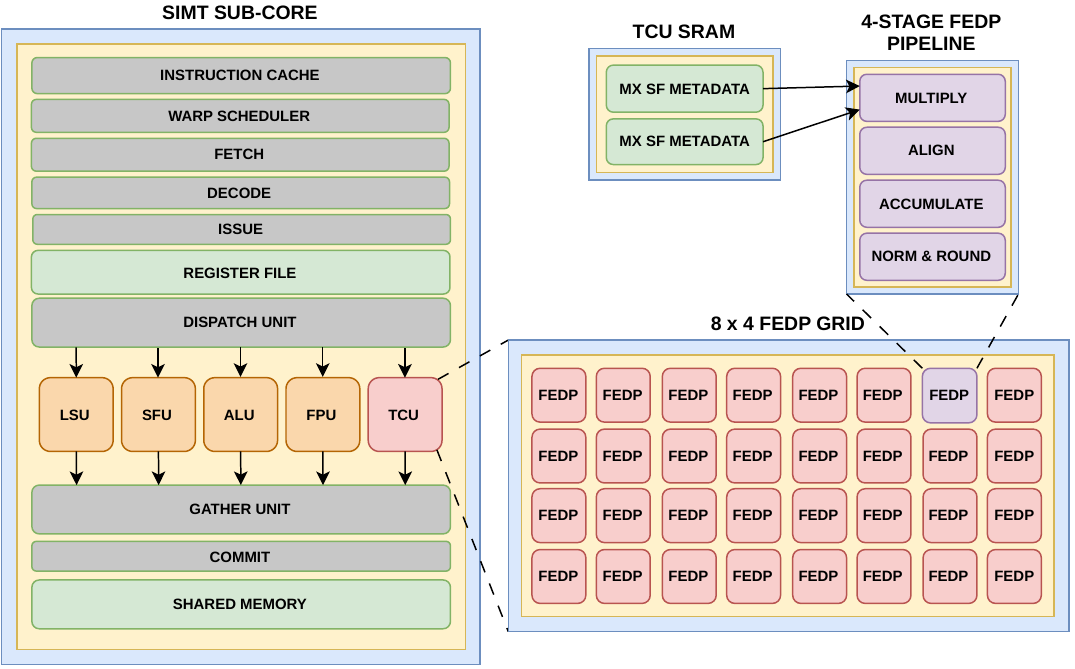}
    \caption{Vortex GPGPU SIMT Sub-Core with the Ten-Four FEDP Microarchitecture based Tensor Core Unit Extension}
    \label{fig:simt_sub_core}
\end{figure}

To address the lack of a high-performance Tensor Core implementation in the open-source GPGPU design space, we introduce \textbf{Ten-Four}, 
a novel configurable mixed-precision Fused Dot Product (FEDP) unit microarchitecture developed as part of the RISC-V Vortex GPGPU's~\cite{10.1145/3466752.3480128} Tensor Core Unit (TCU) Extension. Vortex's configurability at multiple levels of granularity (cores, issue-width, warps, threads) and mature runtime ecosystem present an ideal platform for building and evaluating our design. Fig.~\ref{fig:simt_sub_core} illustrates how we adopt an \texttt{[8x4]} grid of FEDP units to form a TCU within a Vortex GPGPU SIMT sub-core in a 32-threads/warp configuration.

To the best of our knowledge, this is the first major open-source work bridging the gap between specialized Fused Dot Product implementations and SoftGPU Tensor Core prototypes. The key contributions of our work can be summarized as follows:
\begin{itemize}
    \item We propose a configurable 4-cycle Fused Dot Product (FEDP) pipeline supporting low-precision multiplication in FP16, BF16, TF32, FP8(E4M3), BF8(E5M2) and FP4 with higher-precision FP32 accumulation as part of the Vortex TCU.
    \item We fuse the integer and floating-point datapaths through a novel addend-splitting strategy and introduce sparse lane clock-gating to reduce dynamic power consumption while incurring minimal overhead and maximizing resource reuse.
    \item We integrate OCP Microscaling (MX) format support while preserving early addend accumulation using a per-lane scale-factor compensation scheme.
    \item We demonstrate 2.7$\times$-7.9$\times$ speedup over equivalent Berkeley HardFloat and FPnew based implementations at less than 60\% the area cost on the AMD Alveo U55C FPGA
    \item ASIC synthesis using the ASAP7 7nm PDK achieves up to 33.8 TFLOPS/mm$^2$ MXFP8 compute density at 1.5 GHz $F_{\text{max}}$
    \item We prototype LNSF4, a custom NVFP4-variant with LNS8 scale-factors that matches its perplexity with 6.7\% fewer LUTs and 11.1\% lower dynamic power.
\end{itemize}
\section{Background and Motivation}

\begin{table*}[t]
\centering
\caption{Comparison of Prior Floating-Point Arithmetic Libraries and Fused Dot Product Designs}
\label{tab:related_work}
\begin{minipage}{\textwidth}
\renewcommand{\arraystretch}{1.2}
\setlength{\tabcolsep}{3pt}

\resizebox{\textwidth}{!}{%
\begin{tabular}{@{}lcccccccccccc@{}}
\toprule
\multirow{2}{*}{\textbf{Design}}
& \multirow{2}{*}{\textbf{Open-Source}}
& \multicolumn{6}{c}{\textbf{Supported Input Formats}}
& \multirow{2}{*}{\textbf{Configurable}}
& \multirow{2}{*}{\makecell{\textbf{Fused Integer}\\\textbf{Datapath}}}
& \multirow{2}{*}{\textbf{Microscaling}}
& \multirow{2}{*}{\makecell{\textbf{Sparse Lane}\\\textbf{Clock-Gating}}}
\\

\cmidrule(lr){3-8}

&
& \textbf{TF32}
& \textbf{FP16}
& \textbf{BF16}
& \textbf{FP8}
& \textbf{BF8}
& \textbf{FP4}
&
&
&
&
\\

\midrule

Berkeley HardFloat~\cite{hauser2019hardfloat}
& \cmark
& \xmark & \cmark & \xmark & \xmark & \xmark & \xmark
& \cmark
& \xmark & \xmark & \xmark
\\

FPnew~\cite{mach2020fpnew}
& \cmark
& \xmark & \cmark & \cmark & \cmark & \cmark & \xmark
& \cmark & \xmark & \xmark & \cmark
\\

FloPoCo~\cite{ASA2024}
& \cmark
& \xmark & \cmark & \xmark & \xmark & \xmark & \xmark
& \xmark & \xmark & \xmark & \xmark
\\

\midrule

ExSdotp~\cite{9974223}
& \cmark
& \xmark & \cmark & \cmark & \cmark & \cmark & \xmark
& \cmark & \xmark & \xmark & \xmark
\\

MXDOTP~\cite{11113623}
& \cmark
& \xmark & \xmark & \xmark & \cmark & \cmark & \xmark
& \xmark & \xmark & \cmark & \xmark
\\

Desrentes~\emph{et~al.}~\cite{10456822}
& \cmark
& \xmark & \xmark & \xmark & \cmark & \cmark & \xmark
& \xmark & \xmark & \xmark & \xmark
\\

Lutz~\emph{et~al.}~\cite{10579354, 11315161}
& \xmark
& \xmark & \xmark & \xmark & \cmark & \cmark & \xmark
& \xmark & \xmark & \cmark & \xmark
\\

Cuyckens~\emph{et~al.}~\cite{11261796}
& \xmark
& \xmark & \xmark & \xmark & \cmark & \cmark & \cmark
& \xmark & \cmark & \cmark & \xmark
\\

\midrule

\textbf{Ten-Four (This Work)}
& \cmark
& \cmark & \cmark & \cmark & \cmark & \cmark & \cmark
& \cmark
& \cmark & \cmark & \cmark
\\

\bottomrule
\end{tabular}%
}
\end{minipage}
\end{table*}

\subsection{FEDP Computation in Tensor Cores}

Tensor Core WMMA instructions operate on \emph{warp-registers} that cooperatively load and store sub-matrix operands across all 32 threads of a warp. This can be conceptualized as a single 32-bit register (e.g., R0) replicated across all 32 threads, providing $32 \times 32 = 1024$ bits of total operand capacity. Since we can pack two 16-bit, four 8-bit, or eight 4-bit elements per 32-bit register
, assuming FP16 inputs, a single warp-register can hold \texttt{1024/16=64} elements, and yields a \texttt{MxK=8x8} tile for matrix~\texttt{A}. However, since FP32 accumulation limits matrix~\texttt{C} to \texttt{1024/32=32} elements, we must set \texttt{N=4}, finally giving us an \texttt{8x4x8} FP16/FP32 MMA shape. Similarly, with FP8 inputs, the doubled packing density extends the reduction dimension, and produces an \texttt{8x4x16} FP8/FP32 shape instead. 

Thus, a \emph{Tensor Core} can be realized as an \texttt{MxN} grid of \texttt{K}-element FEDP unit instantiations:
\begin{equation}
D_{m,n}^{\textit{FP32}} =
\sum_{i=0}^{K-1}
\left(A_{m,i}^{\textit{FP16}} \times B_{i,n}^{\textit{FP16}}\right)
+ C_{m,n}^{\textit{FP32}}
\end{equation}
Constrained by Vortex's RISC-V ISA, each thread can have up to 32 registers per thread. Allocating this across \texttt{A(8)}, \texttt{B(8)} and \texttt{C/D(8)}, the largest tile each warp can load/store is $\sqrt{(8\times32)}$=\texttt{16x16x16}. Hence, a micro-op sequencer in the Issue stage of the SIMT Sub-Core must iterate over the \texttt{8x4x8} physical Tensor Core in the \texttt{M(2x)}, \texttt{N(4x)} and \texttt{K(2x)} dimension  to complete a WMMA computation.

\subsection{Limitations of Prior Art}

\subsubsection{\textbf{Floating-point libraries.}}
Berkeley HardFloat~\cite{hauser2019hardfloat} is a library for IEEE 754-compliant HDL modules with configurable exponent and significand widths using a recoded intermediate representation. It has been widely adopted in open-source accelerator projects such as Gemmini~\cite{gemmini-dac}, Virgo~\cite{10.1145/3676641.3716281}, Ventus~\cite{10818098}, and the Chipyard SoC Framework~\cite{9099108}. FPnew~\cite{mach2020fpnew} is another parameterized 
transprecision FPU targeting RISC-V F-extension support, with configurable 
formats, pipeline depths, and energy-proportional execution. It has seen usage in the PULP Platform~\cite{7477325}, Vortex FPU 
lanes~\cite{10.1145/3466752.3480128}, and Nada~\emph{et~al.}'s Tensor Core 
prototype~\cite{10946704}. In contrast, FloPoCo~\cite{ASA2024} specifically targets FPGA arithmetic 
generation via parameterizable VHDL operators and automated pipelining.

\subsubsection{\textbf{Specialized fused dot product designs.}}
Multiple foundational works~\cite{7416176,8626475,9974223, 8877403, 6545890, 4674343, ferrere2026vectorized} address the latency, rounding, 
and resource inefficiencies in composing dot product circuits from discrete arithmetic units, and have incorporated features such as shared integer datapaths~\cite{10.1145/2086696.2086720, 11214049}, Posit format support~\cite{10456822,10182007}, Microscaling \cite{10579354,11113623,11261796, 11132989}, and their integration into compute clusters~\cite{10609643,11043957,11539385,11420756, 9154492}.  However, these works remain narrowly scoped to specific formats or ISAs. Ten-Four bridges this gap by taking inspiration from and generalizing these fused microarchitectures into a broader, configurable, mixed-precision GPGPU Tensor Core context, much like the libraries do for 2-operand FPUs, as is summarized in Table~\ref{tab:related_work}.

\subsubsection{\textbf{SoftGPU Tensor Cores on FPGAs.}}
At the same time, prior FPGA work on low-precision AI arithmetic has predominantly only been in custom MAC units for minifloat and MX formats \cite{4762363, 10.1145/3289602.3293912, 10705505, 10705489, 10777058, 11552709, yu2026xtramacefficientmacarchitecture, 10.1145/3695053.3731057, 10946729}, and systolic-array matrix engines \cite{10.1145/3431920.3439282, 10.1145/3773041, 10.1145/3706628.3708867, wang2026mfsamultiformatsystolicarray} targeting standalone dataflow accelerators. However, these designs cannot be plugged in to the current vastly growing open-source SoftGPU on FPGAs ecosystem \cite{10.1145/3466752.3480128, 10818098,  VentusTVLSI2025, 10818116, 10.1145/3676641.3716281, bonet2025riscvmulticoregpusoc, machetti2026egpuopensourceconfigurableriscv} easily without major modifications. Ten-Four addresses this by presenting a configurable FEDP-based Tensor Core engine with tunable knobs for LUT vs DSP-based Wallace-tree multipliers, parameterizable pipeline depths and routing efficiency on the FPGA fabric.
\section{Ten-Four Microarchitecture}


Ten-Four is a high performance, open-source Fused Dot Product Hardware IP~\footnote{
\color{blue} 
https://github.com/vortexgpgpu/vortex/tree/master/hw/rtl/tcu/tfr
} for developing feature-rich mixed-precision GPGPU Tensor Cores, written in SystemVerilog.

\subsection{Key Arithmetic Submodules}
A mixed-precision fused inner dot product datapath requires several multi-operand additions. Carry-Save Adders (CSAs) are particularly suitable for this task, as they effectively reduce N operands of W-bits each to a (W + $\log_2 $N)-bit sum and carry without carry propagation dependencies. We develop a standard CSA using recursively chained 4:2 compressors with conditional 3:2 compressors for odd number of operands cases, alongside a MOD-4 operand grouping CSA to minimize the critical path further when seven or more operands need to be accumulated. The final summation is performed by a Kogge-Stone Adder (KSA), which outperforms carry-lookahead designs by sacrificing area efficiency to achieve lower fanout at every stage through its parallel prefix tree structure. We also implement Wallace tree multipliers (WTMUL) whose partial products are effectively reduced using CSAs. We decide against incorporating Radix-4 Booth recoding in our design, since the incurred bit-pair encoding overhead outweighed the benefit of halving partial products at our target 4-11 bit widths.

\begin{figure}
    \centering
    \includegraphics[width=\linewidth]{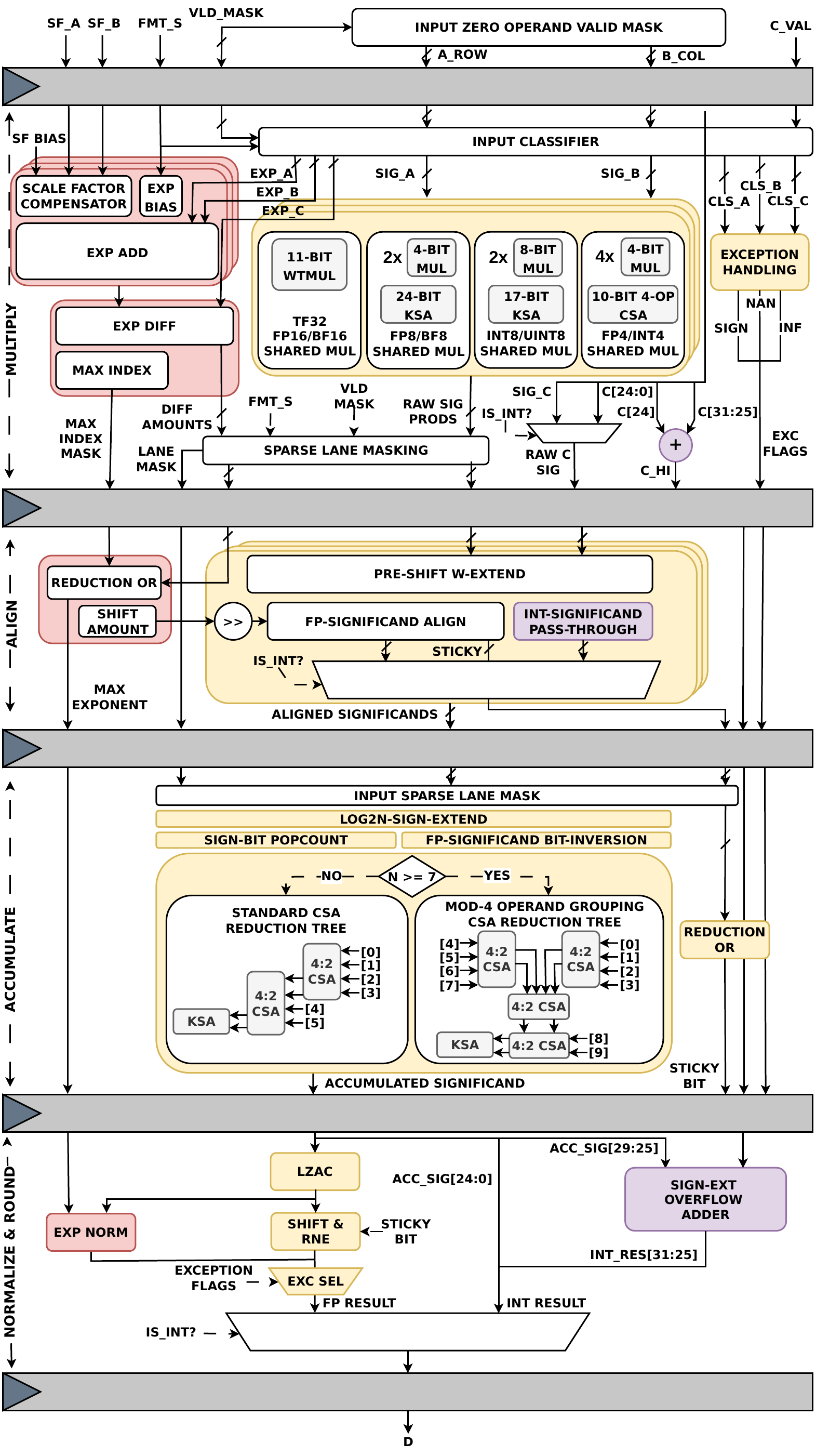}
    \caption{Ten-Four Fused Dot Product Microarchitecture}
    \label{fig:tfr_datapath}
\end{figure}

\subsection{Ten-Four Mixed-Precision Floating-Point Datapath}
Depending on the SIMT sub-core configuration, Ten-Four performs either a four-element or eight-element FEDP at 4/8 or 16/32 threads/warp respectively. It can also be configured to selectively instantiate any subset or all of the available input formats at compile-time to reduce area and improve efficiency depending on the application requirements. For example, LLM training workloads that exclusively use FP16, BF16 or TF32 formats can benefit from omitting the lower-precision format logic entirely. On the other hand, edge devices may only want to instantiate the 4-bit and 8-bit datatypes' circuitry. Ten-Four supports dynamic selection of the input datatype at runtime through the source format signal. As illustrated in Fig.~\ref{fig:tfr_datapath}, the complete Ten-Four datapath consists of 4 pipeline stages: 

\subsubsection{\textbf{Stage-1: Shared Multiplier, Difference Matrix and Exception Handling}}
Ten-Four utilizes a class-wise shared multiplier scheme between formats of similar mantissa bit-widths, making an equal tradeoff between the short critical path of format-dedicated multipliers and the area efficiency of unified sub-word partial product grids as proposed by Zhang~\emph{et~al.}~\cite{8626475} and Ferrere~\emph{et~al.}~\cite{ferrere2026vectorized}. For example, all three FP16, BF16, and TF32 mantissa multiplications are performed in a single 11$\times$11-bit WTMUL. Accordingly, BF16 mantissas need to be zero-extended in the front before entering the shared multiplier to maintain the same input operand width. Remember that, since the inputs are packed as FP16/BF16 pairs or FP8/BF8 quads per 32-bit register, for a given K-element dot product, we instantiate 2$\times$K multiplier lanes in parallel. Since we can only pack one TF32 operand per 32-bit register this way, only every alternate lane is valid for this input type. In contrast, since we can pack four FP8/BF8 elements together, they require an additional reduction per lane to maintain operand count consistency later in the pipeline. Hence, FP8 (E4M3) and BF8 (E5M2) share two 4$\times$4-bit WTMULs whose products are then summed with a 24-bit KSA. In this manner, all formats converge to a raw E8M25 intermediate representation to maintain consistency and efficient resource utilization in subsequent pipeline stages.

Similarly, the exponent sum of the incoming A and B operands, along with their respective format-wise FP32 bias conversions, is computed according to the equations:
\begin{equation}
CONV_{FP16/FP32} = BIAS_{FP32} - (2\times BIAS_{FP16}) + 1
\end{equation}
\begin{equation}
EXP_{FP32} = EXP_{A} + EXP_{B} + CONV_{FP16/FP32}
\end{equation}
These exponent sums are fed into the first part of our maximum exponent identification circuitry, which extends Sohn~\emph{et~al.}'s subtractor-based comparator architecture~\cite{7416176} to support an arbitrary number of $N$ operands. First, all $(N-1)\times(N-1)$ pairwise exponent differences are computed in parallel to form the \emph{difference matrix}. The sign bit of each subtractor result indicates the relative magnitude of its operand pair. To reduce area overhead, we exploit symmetry by computing only the upper triangle and deriving the lower triangle by complementing the sign bits. The maximum exponent index is then extracted from the sign matrix via reduction AND and NOR logic, and passed to the next stage as a one-hot encoded mask.

Additionally, standard IEEE-754 compliant exception handling is performed in parallel to the exponent and mantissa processing. For each product, we detect NaN inputs and infinity-times-zero conditions for multiplication exceptions. Addition exceptions are detected by identifying opposing-signed infinities across the dot product elements and addend. This produces the final result's sign, NaN, and infinity flags preemptively in the very first stage.

\subsubsection{\textbf{Stage-2: Max Exponent, Shift Amount and Significand Alignment}}

This stage completes the max exponent identification via a reduction OR on the one-hot encoded mask from the previous stage. Shift amounts are then derived by reusing the difference matrix and negating values as necessary, yielding near-$O(1)$ critical path complexity (bounded in practice by fanout at higher $N$) at an $O(N^2)$ area cost compared to traditional reduction tree comparators. Finally, the significand products are aligned using the computed shift amounts, with two extra alignment bits and per-lane sticky bits retained from the shifted-out ones to preserve precision.

 \subsubsection{\textbf{Stage-3: Accumulation}}
Naive dot product implementations accumulate the addend "C" separately after completing dot product summation, requiring additional 2-operand alignment, normalization, and rounding that increases both rounding error and critical path delay. Our design instead integrates addend processing from the first pipeline stage, where C's exponent participates in maximum exponent finding and its significand undergoes alignment alongside product terms. 

Here, the aligned significands and addend are first sign-extended to $25 + \log_2(2K)$ bits to handle signed arithmetic correctly. Now, instead of performing a naive 2s complement on each negative term individually before the multi-operand CSA, we only bit-invert the negative significands and defer the $+1$ correction to be collectively handled as an extra operand in the CSA. This value is simply the total number of negative terms in the accumulation, and can be calculated by performing a simple popcount on the sign bits of each lane. Depending on the total operand count -- six for 4/8 threads/warp ($K$ products, one addend, one correction term) or ten for 16/32 threads/warp configurations -- either a standard CSA tree or a MOD-4 operand grouping CSA is selected at compile-time. Per-lane sticky bits are simultaneously reduction-OR'd into a single sticky bit for rounding the final FEDP result.

\subsubsection{\textbf{Stage-4: Normalization and Rounding}}
The final pipeline stage extracts the magnitude from the signed accumulation result and utilizes a predictive Leading Zero Counter (LZAC) for determining the normalization shift amount. The exponent is adjusted by subtracting the computed shift amount from the maximum exponent, while the mantissa is normalized by left-shifting the significand. Standard Round-to-Nearest-Even (RNE) rounding is applied using the LSB, Guard, Round, and previously extracted Sticky bit to produce the final FP32 dot product result. If an exception occurred, the result is overridden with the IEEE-754-compliant canonical NaN or infinity representation.

\subsection{Fusing the Integer Datapath}
Integer dot product operations require multiple arithmetic components already present in the floating-point datapath~\cite{10.1145/2086696.2086720, 11261796}. Fusing both pipelines eliminates the need for an arbiter and scheduling two separate Tensor Core execution units. Hence, Ten-Four makes an effort to do the same by supporting INT8, UINT8, INT4, and UINT4 multiplication with INT32 accumulation within the existing floating-point datapath, by only adding minimal multiplexing overhead. While integer formats have their own class-wise shared multipliers, and simply pass-through stage 2, the significantly more expensive stage 3 accumulator is reused for integer formats as well.

However, fusing the 32-bit integer addend C addition presents a challenge as it exceeds the accumulator width of 25+$\log_2$(2K) bits. We employ a novel splitting strategy to address this constraint. The lower 25 bits of C are accumulated alongside the product terms in the stage 3 CSA, while only the upper 7 bits (denoted C\_HI in Fig.~\ref{fig:tfr_datapath}) propagate through the pipeline, considerably reducing intermediate pipeline register overhead. In the final stage, the upper 7 bits of the integer result are constructed in parallel with floating-point normalization by adding the sign-extended overflow from the accumulator to C\_HI. This is simply concatenated with the lower 25-bit accumulation result to produce the complete INT32 output.

\subsection{Doubling Dot Product Length (FEDP-2K)}
\label{sec:fedp2k}

\begin{figure}
    \centering

    \subfloat[FEDP-K vs. FEDP-2K Tensor Core Tiling and Uop Iterations to Complete one 32-thread WMMA \texttt{16x16x16} Tile Computation
    \label{fig:fedp2k_a}]{
        \includegraphics[width=0.9\columnwidth]{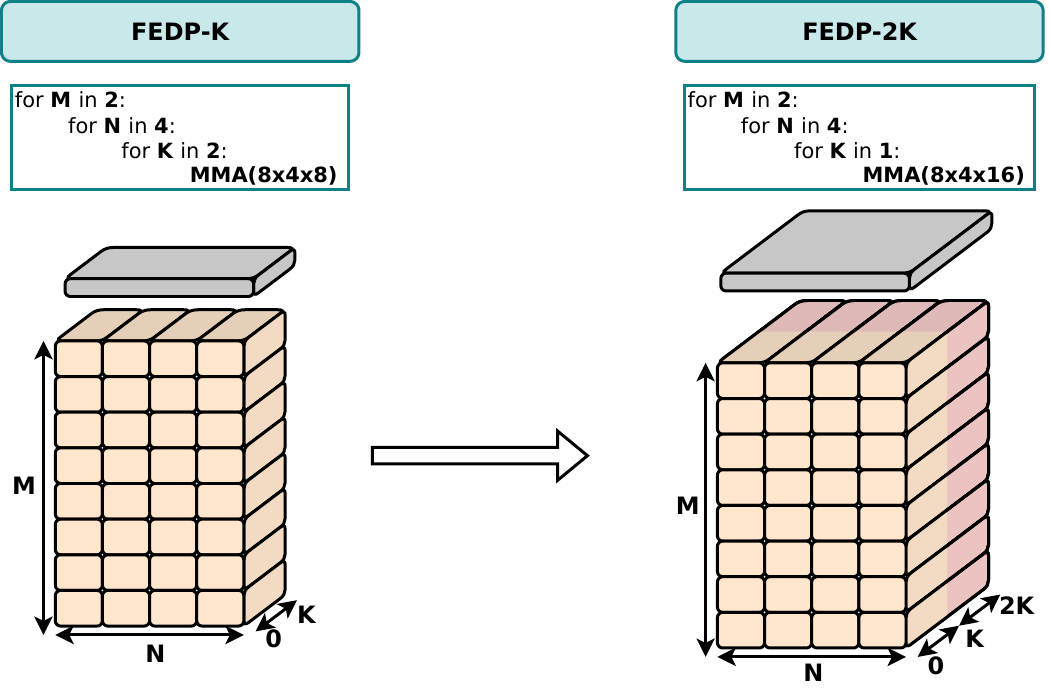}
    }

    \vspace{4pt}

    \subfloat[FEDP-K vs. FEDP-2K Dot Product Length Expansion\label{fig:fedp2k_b}]{
        \includegraphics[width=0.9\columnwidth]{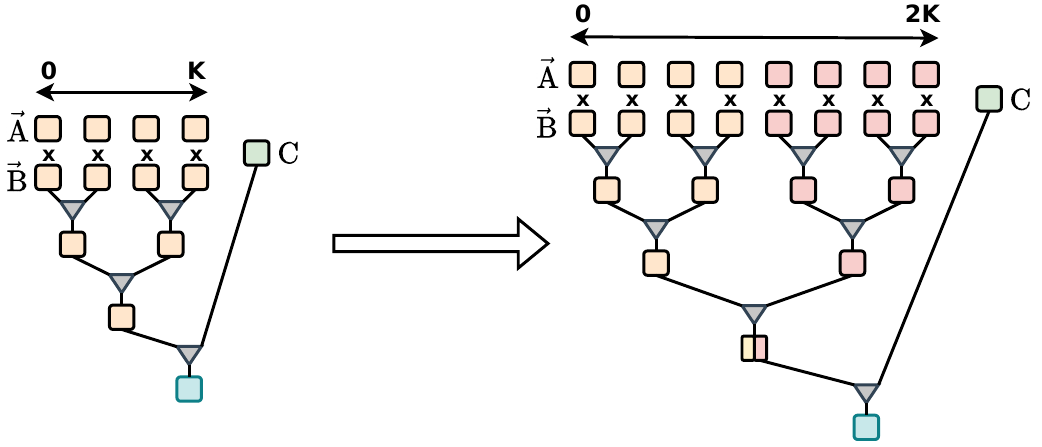}
    }

    \caption{FEDP-2K Microarchitecture extension}
    \label{fig:fedp2k}
\end{figure}

The dot product length $K$ of each FEDP was originally bounded by how much
operand data the GPU Register File could stage. As GEMMs grew larger and took up
an ever greater share of deep learning kernel runtime, GPU designers began loading
Tensor Core inputs directly from shared memory (SMEM), bypassing the Register File
altogether to relieve some of the already extreme register pressure~\cite{nvidia2022hopper}. Sourcing operands from SMEM lifts this cap on $K$ and allows us to feed a second extended chunk of work to each FEDP. We call this design \emph{FEDP-2K} since it doubles the dot product length to $2K$ and halves the number of $K$-dimension micro-op iterations that the sequencer needs to produce, as shown in
Fig.~\ref{fig:fedp2k}.

\subsubsection{\textbf{Backward Compatibility with WMMA}}
Legacy WMMA instructions must still execute correctly with FEDP-2K enabled,
even though there is no upper operand half for them to source from the
Register File. To solve this, we simply feed zero values in this case so that the existing WMMA micro-op semantics are left unchanged. We suspect NVIDIA takes a similar approach in their Hopper architecture~\cite{nvidia2022hopper}, since legacy WMMA there only reach 62.9\% of theoretical peak throughput on average \cite{luo2025dissectingnvidiahopperarchitecture}. Furthermore, since enabling 2:4
structured sparsity halves the $K$-dimension micro-ops as well, when both features are active the $K$-dimension iteration count is lower bounded to one and the same zero-value-hardwiring keeps the FEDP inputs full.

\subsubsection{\textbf{Pipeline Latency Tradeoff}}
Doubling the FEDP width adds critical path delay in two stages.
MULTIPLY sees higher fanout on the shared operand, and ACCUMULATE must reduce
twice as many products, which adds an extra intermediate reduction row even
with a 4:2 compressor-based CSA tree. However, even with deeper FEDP pipelining (say, 8-cycle latency), FEDP-2K remains a
net throughput win because WMMA/WGMMA micro-ops are already heavily pipelined and can fill the extra stages easily without stalling, especially in
larger SGEMM kernels. 

\subsubsection{\textbf{Why Not Extend Further to 4K or 8K-Wide then?}}
We stop at 2K for three reasons. First, each FEDP becomes far more
expensive, as fanout across the datapath and the size of the intermediate
pipeline register buffers grow sharply. Second, a wider unit sits heavily
underutilized on WMMA and WMMA + 2:4 SPARSE instructions. Third, the additional
gains are marginal and appear only at very large SGEMM dimensions. Widening further would also erode the fundamental programmability that
finer-grained Tensor Cores exist to offer over large rigid systolic
arrays in GPUs. It is what lets them handle irregular,
arbitrary-dimension, and sparse workloads effectively.

\subsection{Sparse Lane Mask and Clock Gating}
\begin{figure}[h]
    \centering

    \subfloat[Sparse-Dense interaction\label{fig:fedp_dstca}]{
        \includegraphics[width=0.47\columnwidth]{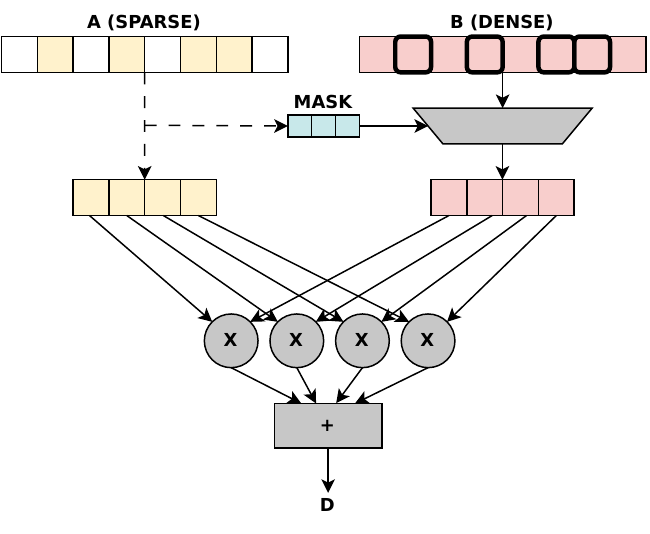}
    }
    \hfill
    \subfloat[Sparse-Sparse interaction\label{fig:fedp_dstcb}]{
        \includegraphics[width=0.47\columnwidth]{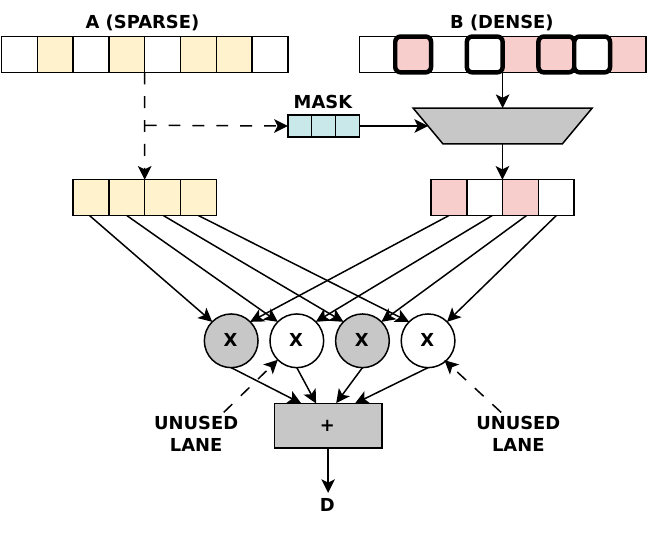}
    }

    \caption{Inner-Product Primitive Dual-Side Unstructured Sparsity Limitations}
    \label{fig:combined_inner_product}
\end{figure}

Modern deep learning workloads, including pruned LLMs, recommendation systems, and graph neural networks, naturally exhibit substantial sparsity in both weights and dynamic activations. While exploiting this dual-side sparsity could significantly reduce memory footprint and bandwidth requirements, current NVIDIA Sparse Tensor Cores~\cite{nvidia2020ampere, nvidia2022ada, nvidia2022hopper, nvidia2024blackwell} only support 2:4 structured sparsity on the weight matrix due to the fundamental unused lane limitation of the inner-product computation primitive itself as illustrated in Fig.~\ref{fig:combined_inner_product}. 

Wang~\emph{et~al.} proposed Dual-Side Sparse Tensor Cores (DSTC)~\cite{dstc9499745} that circumvent this limitation by replacing the inner-dot-product units with outer-dot-product units. Outer-products naturally avoid the inner-join problem by computing cross-products between column-row vector pairs, thereby condensing sparse inputs into dense vectors before multiplication. However, outer-product requires storing full M$\times$N intermediate partial matrices in expensive accumulation buffers across K reduction steps, incurring substantial area overhead and reduced computational density for a given die.

Ten-Four adopts a pragmatic middle ground by retaining the area-efficient inner-product FEDP design while at least leveraging sparsity for power reduction through selective lane clock gating. As shown in Fig.~\ref{fig:tfr_datapath}, an input valid mask derived from the source operand format and zero-detection logic controls the clock gating of FEDP lanes. When a lane's input is identified as zero, its pipeline registers are clock-gated from the very start, eliminating switching activity through the multiply and align stages where lane computations are self-contained. However, before entering the accumulator stage where multi-lane reduction occurs, the third pipeline registers' outputs are AND-gated with the valid lane mask, ensuring disabled lanes provide zeros to the CSA rather than stale register values. This approach reduces dynamic power consumption without the microarchitectural complexity or area overhead of outer-product designs, making it a practical solution for Tensor Cores targeting dual-side sparse workloads. Notably, this scheme is orthogonal to 2:4 structured sparsity (whose speedup and power savings come from halving WMMA micro-ops in the $K$-dimension), and can 
be applied alongside it.

\subsection{Microscaling (MX) Format Support}
Block-floating-point formats were introduced to preserve higher model accuracy over per-tensor 
quantization while retaining the memory and throughput benefits of low-precision 
arithmetic~\cite{rouhani2023microscalingdataformatsdeep, 10.1145/3579371.3589351}. 
Interestingly, the OCP MX specification~\cite{ocp2023mx} leaves the dot product's internal 
precision and operation order as implementation-defined, allowing for aggressive 
microarchitectural flexibility.

The conventional approach to designing a Microscaling FEDP datapath would involve adding the two E8M0 scale factors $X^{(A)}$ and 
$X^{(B)}$, removing FP32 bias ($-254$), and applying them to the sum-of-products (SoP) before final accumulation:
\begin{equation}
D = \left[ X^{(A)} X^{(B)} \sum_{i=1}^{k} \left(A_i \times B_i \right) \right] + C
\end{equation}
However, this deferred scaling is incompatible with Ten-Four's pipeline, since the addend 
$C$ is processed alongside the SoP from Stage~1. Ten-Four resolves this by simply rearranging the factorization of the scale factors as:
\begin{equation}
D = \sum_{i=1}^{k} \left[ X^{(A)} X^{(B)} \left(A_i \times B_i \right) \right] + C
\end{equation}
This per-lane scale-factor compensation scheme, combined with the existing low-precision exponent and 
bias circuitry, enables Ten-Four to support MXFP8, MXBF8, MXFP4, NVFP4 and MXINT8 with bare-minimum 
modifications, as illustrated in Fig.~\ref{fig:tfr_datapath}.

\section{Evaluation}

\subsection{Floating-Point Library Comparison}
We evaluate Ten-Four against equivalent Xilinx DSP IP, Berkeley HardFloat \cite{hauser2019hardfloat} and FPnew \cite{mach2020fpnew} based discrete TCU implementations, targeting 300~MHz operational clock frequency on the AMD Alveo U55C FPGA~\cite{amd_ds978_u55c} across a range of threads/warp configurations (NT = 4, 8, 16, 32).

\begin{figure}[h]
    \centering
    \includegraphics[width=0.9\linewidth]{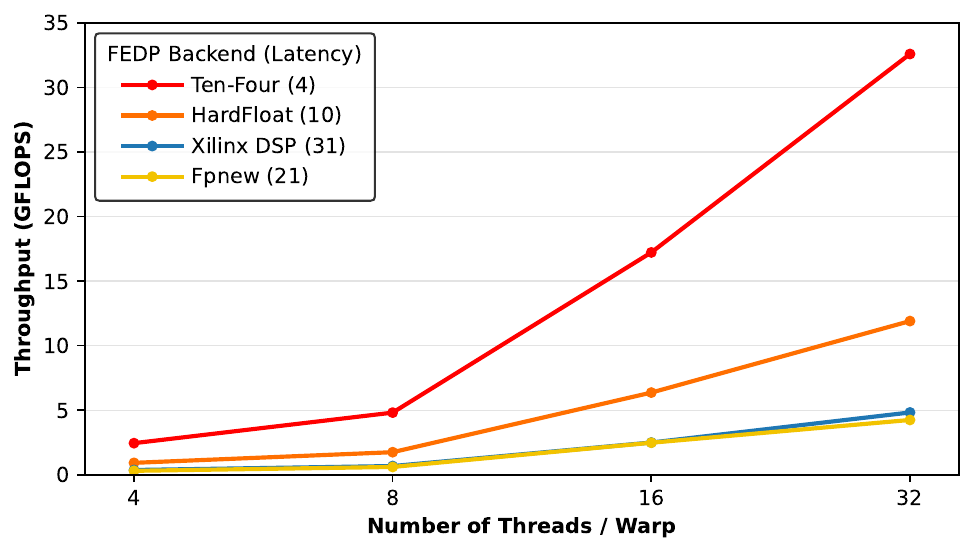}
    \caption{FEDP Backends Performance Scaling (FP16/BF16)}
    \label{fig:Throughput_scaling}
\end{figure}

Fig.~\ref{fig:Throughput_scaling} shows that Ten-Four achieves significantly 
higher single-cycle throughput\footnote{Single-cycle Throughput = (FLOPs / Latency) $\times$ $F_{max}$} scaling (2.446--32.592~GFLOPS) 
compared to Berkeley HardFloat (0.922--11.903 GFLOPS, $\sim$2.7$\times$), Xilinx DSP 
(0.367--4.824 GFLOPS, $\sim$6.7$\times$), and FPnew (0.304--4.244 GFLOPS, 
$\sim$7.9$\times$). This improvement largely stems from our 4-cycle latency and MOD-4 grouping CSA.

\begin{figure}[h]
    \centering
    \includegraphics[width=0.9\linewidth]{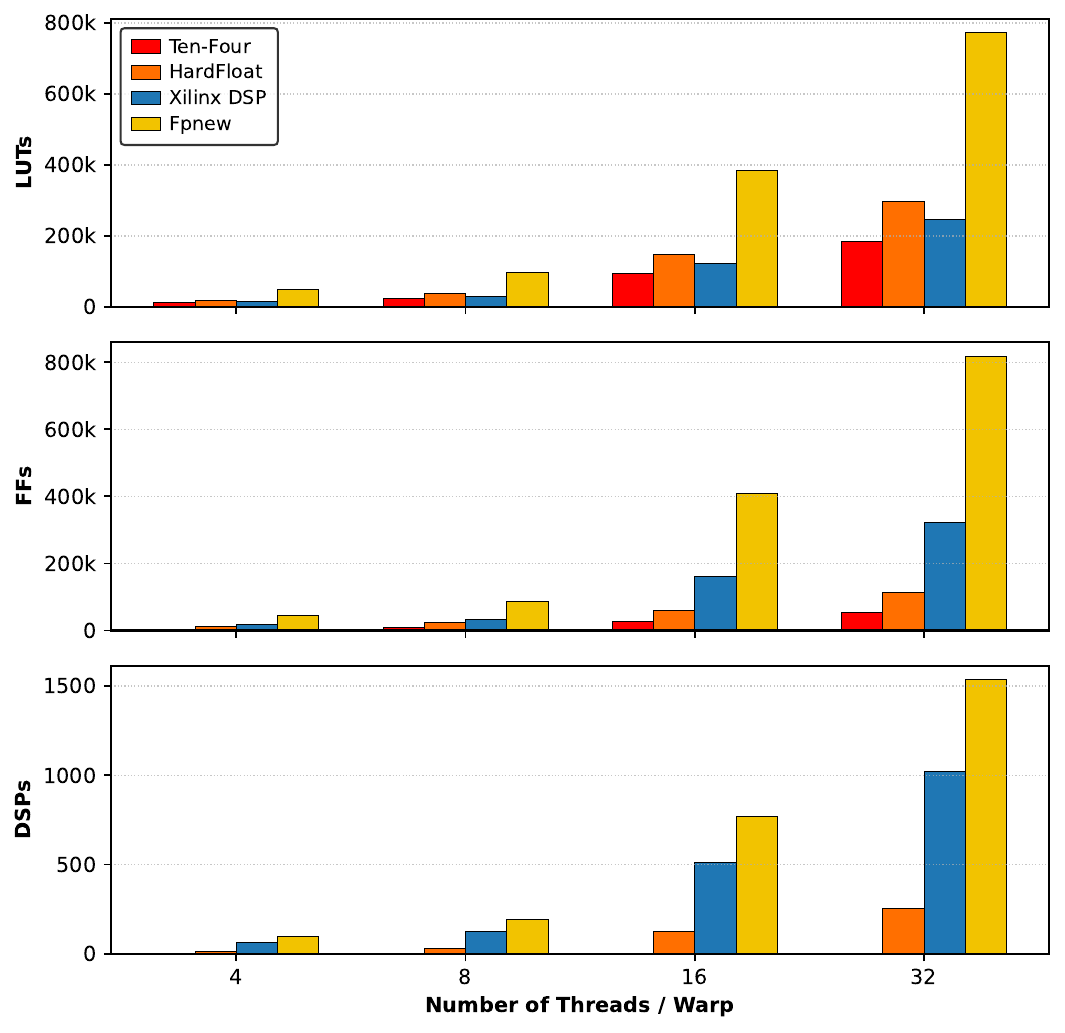}
    \caption{FEDP Backends Post-Implementation Area Utilization (FP16/BF16)}
    \label{fig:fpga_area}
\end{figure}

Furthermore, as demonstrated in Fig.~\ref{fig:fpga_area}, Ten-Four achieves 37--38\% LUT reduction versus HardFloat, 24--25\% versus Xilinx DSP, and 75--76\% versus FPnew, despite eliminating DSP block usage entirely in the LUT-based WTMUL configuration. Flip-Flop usage is also reduced by 52--66\% compared to HardFloat, 75--83\% compared to Xilinx DSP, and 90--93\% compared to FPnew.

\subsection{Class-wise vs Format-Specific WTMULs}

Our proposed class-wise shared multiplier scheme achieves a $\sim$39--40\% reduction in LUT utilization over the format-specific variant as shown in Table~\ref{tab:mult_sharing}. We further prototyped this design using U55C DSP blocks in place of the LUT-based WTMULs to improve FPGA resource utilization when instantiating a full Vortex GPGPU core, and observed a further drop in LUT count of $\sim$32--37\%, at the cost of 21 additional DSP blocks per lane.

\begin{table}[h]
\centering
\caption{Comparing Discrete Format-Specific and Class-wise Shared Multiplier Schemes}
\label{tab:mult_sharing}
\small
\setlength{\tabcolsep}{2.5pt}
\renewcommand{\arraystretch}{1.12}

\begin{tabular}{@{}cccccccc@{}}
\toprule
\textbf{Backend} &
\textbf{N} &
\textbf{SF} &
\textbf{Sharing} &
\textbf{LUT} &
\textbf{FF} &
\textbf{DSP} &
\makecell{\textbf{Dynamic}\\\textbf{Power (mW)}} \\
\midrule

\multirow{4}{*}{Wallace}
& \multirow{2}{*}{4} & \multirow{2}{*}{1}
& Discrete   & 10560 & 415 & 0 & 127 \\
& & & Class-wise & 6452 & 474 & 0 & 216 \\
& \multirow{2}{*}{8} & \multirow{2}{*}{2}
& Discrete   & 21578 & 719 & 0 & 263 \\
& & & Class-wise & 13053 & 715 & 0 & 446 \\

\midrule

\multirow{4}{*}{DSP}
& \multirow{2}{*}{4} & \multirow{2}{*}{1}
& Discrete   & 5106 & 404 & 84 & 131 \\
& & & Class-wise & 4391 & 404 & 42 & 197 \\
& \multirow{2}{*}{8} & \multirow{2}{*}{2}
& Discrete   & 9723 & 715 & 168 & 249 \\
& & & Class-wise & 8286 & 715 & 84 & 373 \\

\bottomrule
\end{tabular}
\end{table}

\subsection{Dual-Side Sparse Lane Mask Power Analysis}

Next, we evaluate the sparse lane mask clock-gating scheme by sweeping dual-side 
sparsity from 0--100\% in 10\% steps across all supported formats on a 
\texttt{64x64x64} SGEMM kernel, using post-implementation SAIF-based power 
estimation on the U55C FPGA. While Static power remains nearly constant 
($\sim$3.30--3.39~W) across all formats and sparsity levels, FP16 exhibits the highest dynamic power consumption ($\sim$5.2~W), while TF32 and BF16 draw slightly less power due to their alternate-lanes-active-only scheme and smaller mantissa bit-width respectively. FP8/BF8 and integer formats consistently consume the least due to their leaner multiplier logic. Notably, however, INT4 can exceed INT8 dynamic power beyond 70\% sparsity levels, because its finer-granularity lane mask introduces proportionally higher clock-gating control logic overhead relative to the reduced compute activity.

\begin{figure}[h]
    \centering
    \includegraphics[width=\linewidth]{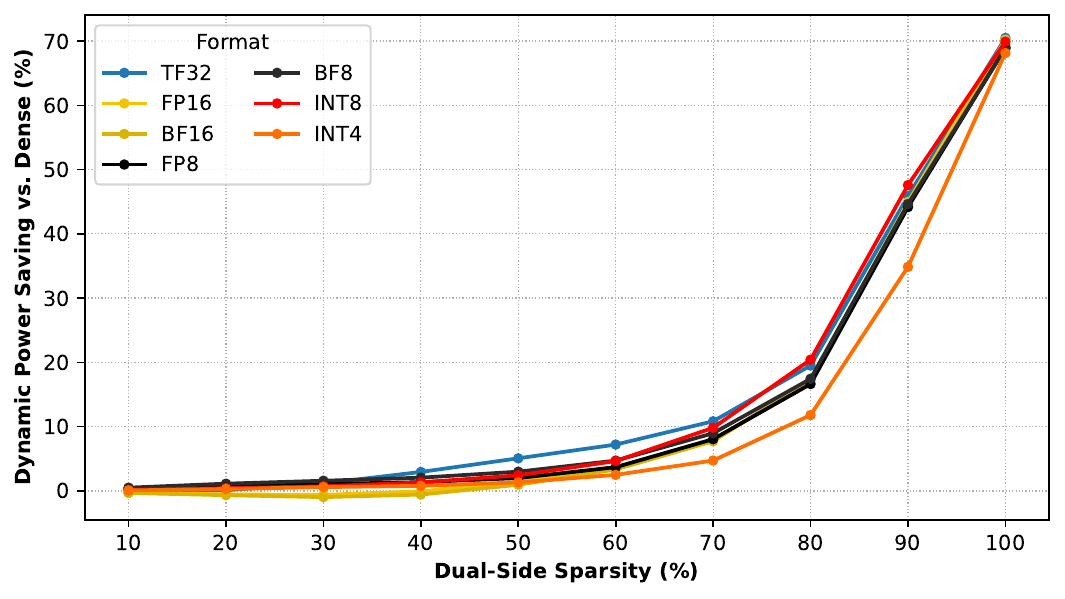}
    \caption{Dynamic Power Saving over Dense at Varying Dual-Side Sparsity levels}
    \label{fig:dsm_pct_saving}
\end{figure}

Fig.~\ref{fig:dsm_pct_saving} demonstrates the corresponding dynamic power savings over the dense baseline. The sparse lane mask becomes increasingly compelling beyond 70\% dual-side sparsity, improving from 4.7--10.8\% savings at 70\% to 34.9--47.6\% at 90\% and 68.1--70.5\% at 100\%. These results show that sparse lane clock-gating is an effective mechanism that can be coupled with existing 2:4 structured sparsity schemes to exploit dynamic power reduction in dual-side sparse workloads, with bare-minimum overhead.

\begin{figure}[h]
    \centering
    \includegraphics[width=\linewidth]{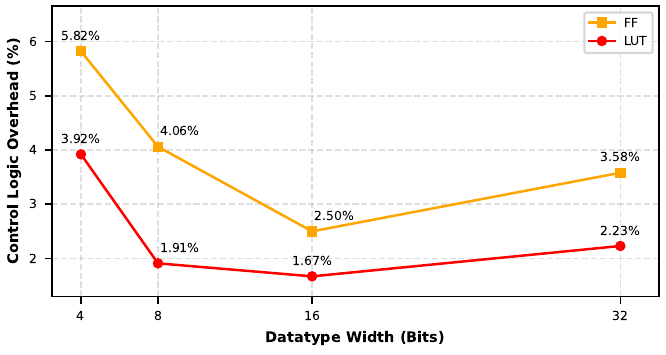}
    \caption{Sparse Lane Mask Clock-Gating Control Logic Overhead \% across Low-precision Format-Group Bit-Widths}
    \label{fig:dsm_overhead}
\end{figure}

As shown in Fig.~\ref{fig:dsm_overhead}, clock-gating control logic overhead in LUTs and FFs increases nearly exponentially with decreasing datatype width due to our low-precision operand packing scheme within 32-bit registers. The 16-bit datatype group incurs the lowest overhead, as Ten-Four was designed at a 16-bit lane processing granularity. Consequently, since only every alternate lane is active for TF32, it incurs slightly higher overhead from the additional indexing control logic. Alternatively, adopting a 32-bit lane granularity would've shifted the minimum overhead to TF32, but would've required an additional first-stage lane reduction for 16-bit formats and doubled the number of reductions for 4-bit and 8-bit formats, adversely affecting the multiplier stage's critical path.

\subsection{Numerical Accuracy Verification}
To validate Ten-Four's FEDP computation, we develop a verification framework using PyTorch-based CUDA kernel generation to produce format-specific WMMA and PTX routines targeting NVIDIA's Ada Architecture~\cite{nvidia2022ada} RTX~4090 and Blackwell Architecture~\cite{nvidia2024blackwell} RTX~5080 GPUs as the hardware reference. Our verification harness systematically exercises corner cases across six distinct feature classes: normals, subnormals, zeros, infinities, NaNs and catastrophic cancellation scenarios. Ten-Four produces bit-exact results matching the NVIDIA Ada and Blackwell Tensor Core outputs for all supported floating-point (TF32, FP16, BF16, FP8, BF8, MXFP8, MXBF8, MXFP4 and NVFP4), and integer (INT8 and INT4) formats, across all 1,000,000+ randomized test vectors.

We further port the FTTN benchmark~\cite{10.1007/978-3-031-93706-4_21, 10701338} through the Vortex WMMA 
C++ API to verify IEEE-compliant subnormal handling and rounding, and compare Ten-Four's 
precision characteristics against analytical numerical accuracy
models of multiple commercial GPUs~\cite{fasi2021numerical, khattak2025accuratemodelsnvidiatensor}, 
as summarized in Table~\ref{tab:gpu_generation}(a).

\begin{table*}[t]
\centering
\caption{Ten-Four (a) Numerical Behavior and (b) FPGA Resource Utilization as it Scales its Supported Formats and Features (2:4 Structured Sparsity, FEDP-2K and Microscaling) across NVIDIA Tensor Core Generations, Excluding FP64 and INT1}
\label{tab:gpu_generation}
\footnotesize
\setlength{\tabcolsep}{1.5pt}
\renewcommand{\arraystretch}{1.08}
\newcommand{\fmts}[1]{{\renewcommand{\arraystretch}{0.8}\begin{tabular}[c]{@{}l@{}}#1\end{tabular}}}

\begin{tabular*}{\textwidth}{@{\extracolsep{\fill}}lllccccccccccc@{}}
\toprule
& & &
\multicolumn{5}{c}{\textbf{\small (a) Numerical Accuracy}} &
\multicolumn{6}{c}{\textbf{\small (b) FPGA Synthesis}}
\\
\cmidrule(lr){4-8}
\cmidrule(lr){9-14}
\multirow{2}{*}{\textbf{Generation}} &
\multirow{2}{*}{\makecell[l]{\textbf{Enabled}\\\textbf{Formats}}} &
\multirow{2}{*}{\textbf{Design}} &
\multirow{2}{*}{\makecell{\textbf{Subnormal}\\\textbf{In / Out}}} &
\multirow{2}{*}{\makecell{\textbf{Product}\\\textbf{Align Bits}}} &
\multirow{2}{*}{\makecell{\textbf{Accumulator}\\\textbf{Width}}} &
\multirow{2}{*}{\makecell{\boldmath$\mathbf{N}_{\mathrm{FMA}}$\\\textbf{Width}}} &
\multirow{2}{*}{\makecell{\textbf{Rounding}\\\textbf{Mode}}} &
\multirow{2}{*}{\makecell{\textbf{F}\boldmath$_{\mathbf{max}}$\\\textbf{(MHz)}}} &
\multirow{2}{*}{\makecell{\textbf{LUT}\\\textbf{(K)}}} &
\multirow{2}{*}{\makecell{\textbf{LUT}\\\textbf{RAM}}} &
\multirow{2}{*}{\makecell{\textbf{FF}\\\textbf{(K)}}} &
\multirow{2}{*}{\textbf{DSP}} &
\multirow{2}{*}{\makecell{\textbf{Power}\\\textbf{(W)}}}
\\[-1pt]
& & & & & & & & & & & & &
\\
\midrule

\multirow{3}{*}{Volta~\cite{nvidia2017volta}}
& \multirow{3}{*}{\fmts{FP16}}
& V100              & \cmark & $(2,23)$ & 28 & 4  & Trunc.
& \multirow{3}{*}{302.6} & \multirow{3}{*}{53.9} & \multirow{3}{*}{336} & \multirow{3}{*}{21.8} & \multirow{3}{*}{128} & \multirow{3}{*}{1.37}
\\
& & \textbf{Ten-Four} & \cmark & $(2,25)$ & 30 & 8  & RTN-TE & & & & & & \\
& & \textbf{(FEDP-2K)}  & \cmark & $(2,25)$ & 31 & 16 & RTN-TE & & & & & & \\
\midrule

\multirow{3}{*}{Turing~\cite{nvidia2018turing}}
& \multirow{3}{*}{\fmts{FP16, INT8,\\INT4}}
& T4                & \cmark & $(2,24)$ & 29 & 4  & Trunc.
& \multirow{3}{*}{300.7} & \multirow{3}{*}{142.3} & \multirow{3}{*}{336} & \multirow{3}{*}{69.2} & \multirow{3}{*}{1280} & \multirow{3}{*}{5.88}
\\
& & \textbf{Ten-Four} & \cmark & $(2,25)$ & 30 & 8  & RTN-TE & & & & & & \\
& & \textbf{(FEDP-2K)}  & \cmark & $(2,25)$ & 31 & 16 & RTN-TE & & & & & & \\
\midrule

\multirow{3}{*}{Ampere~\cite{nvidia2020ampere}}
& \multirow{3}{*}{\fmts{FP16, BF16,\\TF32, INT8,\\INT4}}
& A100              & \cmark & $(2,24)$ & 29/30 & 4/8  & Trunc.
& \multirow{3}{*}{297.4} & \multirow{3}{*}{154.1} & \multirow{3}{*}{636} & \multirow{3}{*}{69.4} & \multirow{3}{*}{1280} & \multirow{3}{*}{5.85}
\\
& & \textbf{Ten-Four} & \cmark & $(2,25)$ & 30 & 4/8  & RTN-TE & & & & & & \\
& & \textbf{(FEDP-2K)}  & \cmark & $(2,25)$ & 31 & 8/16 & RTN-TE & & & & & & \\
\midrule

\multirow{3}{*}{\makecell[l]{Hopper~\cite{nvidia2022hopper}\\Ada~\cite{nvidia2022ada}}}
& \multirow{3}{*}{\fmts{FP16, BF16, TF32,\\FP8, BF8,\\INT8, INT4}}
& H100              & \cmark & $(2,25)$/$(2,25)$/$(2,13)$ & 31/32/21 & 8/16/32 & Trunc.
& \multirow{3}{*}{242.4} & \multirow{3}{*}{394.0} & \multirow{3}{*}{636} & \multirow{3}{*}{82.2} & \multirow{3}{*}{3072} & \multirow{3}{*}{7.63}
\\
& & \textbf{Ten-Four} & \cmark & $(2,25)$ & 30 & 4/8/16  & RTN-TE & & & & & & \\
& & \textbf{(FEDP-2K)}  & \cmark & $(2,25)$ & 31 & 8/16/32 & RTN-TE & & & & & & \\
\midrule

\multirow{3}{*}{Blackwell~\cite{nvidia2024blackwell}}
& \multirow{3}{*}{\fmts{FP16, BF16, TF32,\\FP8, BF8, INT8,\\MXFP8, MXBF8,\\MXFP4, NVFP4}}
& B200              & \cmark & $(2,25)$ & 31/32/33/34 & 8/16/32/64 & Trunc.
& \multirow{3}{*}{227.0} & \multirow{3}{*}{447.7} & \multirow{3}{*}{2684} & \multirow{3}{*}{88.2} & \multirow{3}{*}{3072} & \multirow{3}{*}{13.74}
\\
& & \textbf{Ten-Four} & \cmark & $(2,25)$ & 30 & 4/8/16/32  & RTN-TE & & & & & & \\
& & \textbf{(FEDP-2K)}  & \cmark & $(2,25)$ & 31 & 8/16/32/64 & RTN-TE & & & & & & \\
\bottomrule
\end{tabular*}
\end{table*}

\subsection{FPGA Placement \& Routing}

\begin{figure}[h]
    \centering

    \subfloat[Cell Placement\label{fig:u55c_pr_a}]{
        \includegraphics[width=0.52\columnwidth]{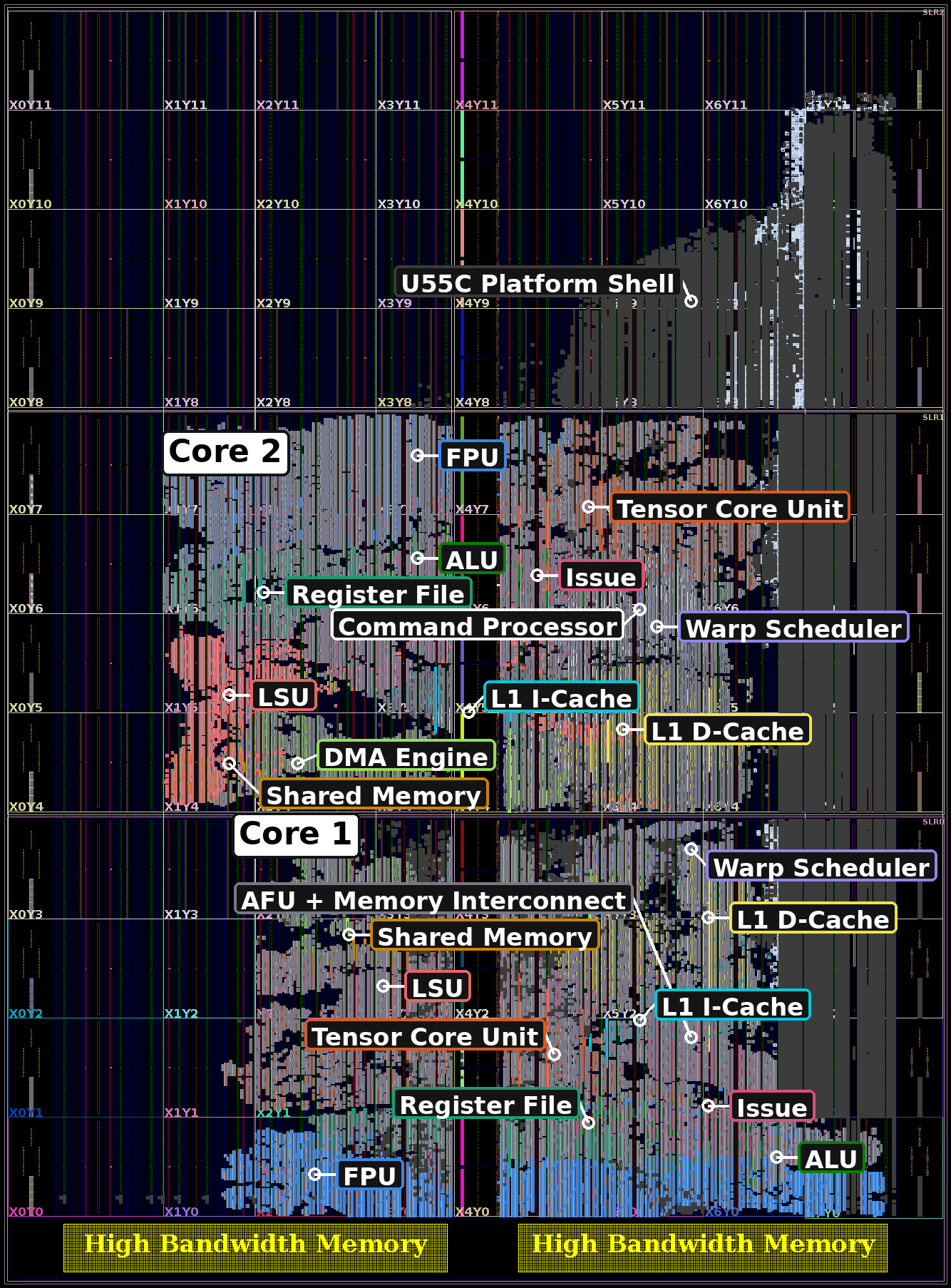}
    }
    \hfill
    \subfloat[Sub-Core Area Breakdown\label{fig:u55c_pr_b}]{
        \includegraphics[width=0.43\columnwidth]{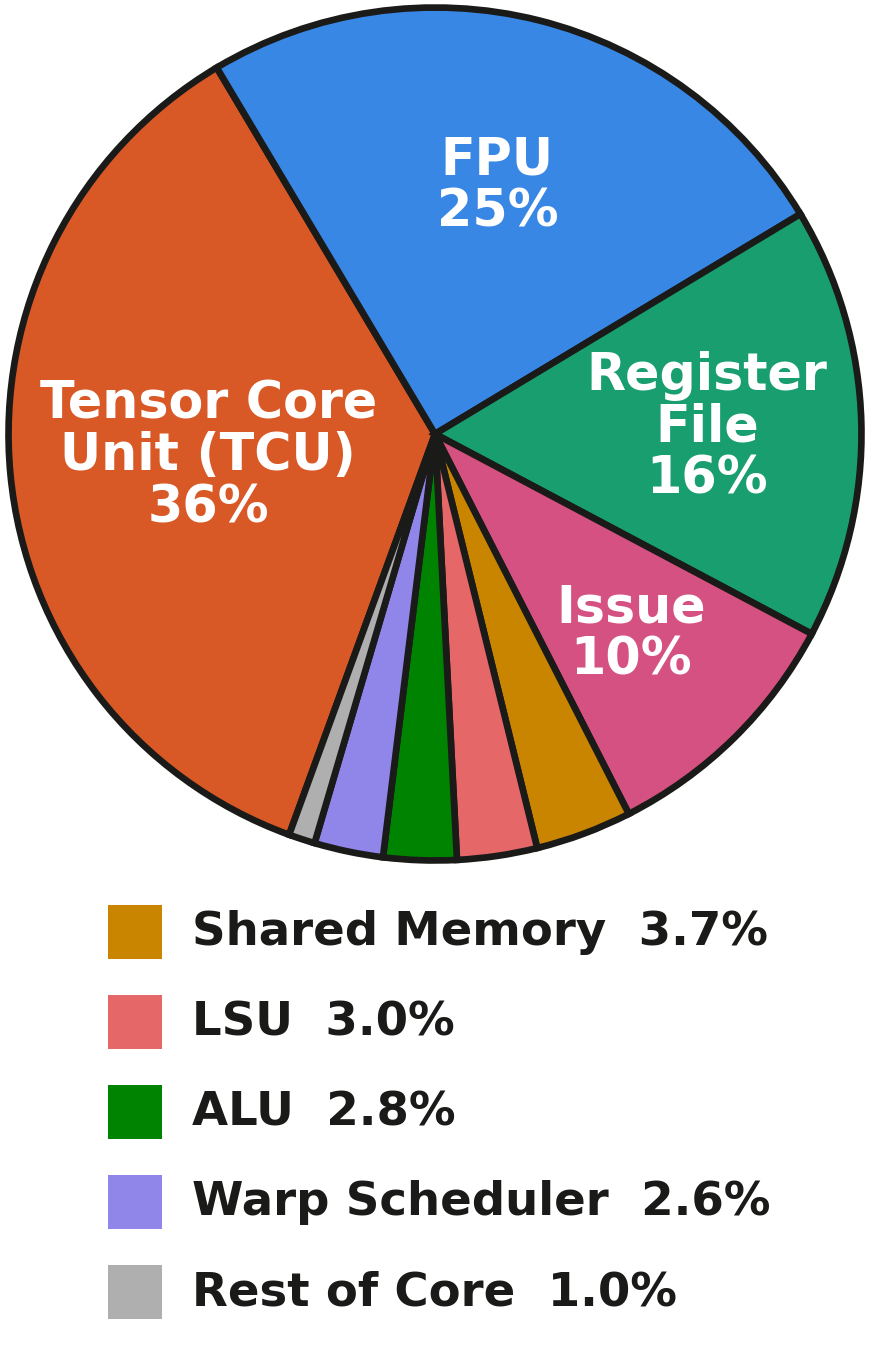}
    }

    \caption{Ten-Four Extended Two-core Vortex GPGPU Instance Placed on an AMD Alveo U55C FPGA}
    \label{fig:u55c_pr}
\end{figure}

Fig.~\ref{fig:u55c_pr_a} shows the placement of a two-core Ten-Four extended Vortex GPGPU instance on the AMD Alveo U55C FPGA~\cite{amd_ds978_u55c}, while Fig.~\ref{fig:u55c_pr_b} breaks down the area contribution of each unit to a single core. The U55C consists of three stacked dies, known as super logic regions (SLRs)~\cite{amd_u55c_platform}. Signals that cross between SLRs travel over a limited pool of super long lines, which add significant delay and clock skew compared to on-die routing. Each core's tightly coupled datapaths, such as the operand collectors feeding the execution units and the DMA engine moving data from global to shared memory, should therefore stay coupled on one die. We use placement blocks to confine each core, along with its private L1 caches and DMA engine, to a single SLR. As a result, only the registered memory-side interface crosses a die boundary. Without this constraint, the place-and-route flow would spread cores across dies, and would fail to route thousands of nets and close timing at a much lower clock frequency. We place the cores in SLR0 and SLR1 rather than SLR1 and SLR2 because the HBM controllers sit along the bottom edge of SLR0. Doing so keeps both cores within one die crossing of global memory, whereas a core in SLR2 would need two crossings to reach HBM. Finally, Table~\ref{tab:gpu_generation}(b) lists the TCU's FPGA resource utilization across advancing NVIDIA generations.

\subsection{ASIC Design Flow Analysis}
We also synthesize Ten-Four in all four-element 
(NT=4,8), eight-element (NT=16,32), and FEDP-2K dot product configurations using Synopsys Design 
Compiler and the ASAP7 7nm Predictive PDK~\cite{CLARK2016105}. Each configuration is 
synthesized in two variants: a minimal MXFP8/MXBF8-only datapath to establish a lean area baseline, and a fully-featured variant with all floating-point and 
integer formats enabled to capture the overhead of format-wise shared logic fanout. All runs target the \texttt{asap7sc7p5t\_AO\_LVT\_TT\_nldm} standard cell library at 1500~MHz 
under \texttt{PVT\_0P7V\_25C} typical conditions. As shown in Table~\ref{tab:dotp_comparison}, Ten-Four outperforms other open-source and research SOTA FP8 fused dot-product microarchitectures while supporting a broad set of additional features and formats. The MXFP8-only variants reach 30.9--33.8 TFLOPS/mm$^2$ compute density, a 5.6--6.1$\times$ gain over MXDOTP, the only prior design with matching $2\times8$b scale support.

\begin{table}
\centering
\caption{ASIC Synthesis Comparison Between FP8 Dot-Product Accelerator Designs}
\label{tab:dotp_comparison}
\footnotesize
\setlength{\tabcolsep}{1.5pt}
\renewcommand{\arraystretch}{1.2}

\begin{tabular}{lcccccc}
\toprule
\textbf{Design} &
\textbf{Tech.} &
\textbf{Voltage} &
\textbf{Freq.} &
\textbf{Area} &
\textbf{Throughput} &
\textbf{MX Scale} \\

&
\textbf{nm} &
\textbf{V} &
\textbf{GHz} &
\textbf{mm$^2$} &
\textbf{GFLOPS} &
\textbf{Support} \\
\midrule

ExSdotp \cite{9974223}
& 12
& 0.8
& 1.26
& $5.13 \times 10^{-3}$
& 20.2
& \xmark \\

Desrentes \textit{et al.} \cite{10456822}
& 16
& ---
& 1.25
& $9.81 \times 10^{-3}$
& 80.0
& \xmark \\

Lutz \textit{et al.} \cite{10579354, 11315161}
& 5
& ---
& 3.60
& $6.74 \times 10^{-4}$
& 28.8
& \cmark\ $(1 \times 7\mathrm{b})$ \\

MXDOTP \cite{11113623}
& 12
& 0.8
& 1.09
& $3.15 \times 10^{-3}$
& 17.4
& \cmark\ $(2 \times 8\mathrm{b})$ \\

\multirow{3}{*}{\textbf{Ten-Four (Ours)}}
& \multirow{3}{*}{\textbf{7}}
& \multirow{3}{*}{0.7}
& \textbf{1.50}
& $\mathbf{7.76 \times 10^{-4}}$
& \textbf{24.0}\makebox[0pt][l]{\textsuperscript{$\dagger$}}
& \cmark\ $\mathbf{(2 \times 8\mathrm{b})}$ \\

&
&
&
\textbf{1.50}
& $\mathbf{1.45 \times 10^{-3}}$
& \textbf{48.0}\makebox[0pt][l]{\textsuperscript{$\ddagger$}}
& \cmark\ $\mathbf{(2 \times 8\mathrm{b})}$ \\

&
&
&
\textbf{1.50}
& $\mathbf{2.84 \times 10^{-3}}$
& \textbf{96.0}\makebox[0pt][l]{\textsuperscript{$\ddagger$}\textsuperscript{$\S$}}
& \cmark\ $\mathbf{(4 \times 8\mathrm{b})}$ \\

\midrule

\multirow{3}{*}{\textbf{Ten-Four (All fmts)}}
& \multirow{3}{*}{\textbf{7}}
& \multirow{3}{*}{0.7}
& \textbf{1.26}
& $\mathbf{1.91 \times 10^{-3}}$
& \textbf{20.1}\makebox[0pt][l]{\textsuperscript{$\dagger$}\textsuperscript{*}}
& \cmark\ $\mathbf{(2 \times 8\mathrm{b})}$ \\

&
&
&
\textbf{1.16}
& $\mathbf{3.58 \times 10^{-3}}$
& \textbf{37.2}\makebox[0pt][l]{\textsuperscript{$\ddagger$}\textsuperscript{*}}
& \cmark\ $\mathbf{(2 \times 8\mathrm{b})}$ \\

&
&
&
\textbf{1.12}
& $\mathbf{7.19 \times 10^{-3}}$
& \textbf{71.9}\makebox[0pt][l]{\textsuperscript{$\ddagger$}\textsuperscript{$\S$}\textsuperscript{*}}
& \cmark\ $\mathbf{(4 \times 8\mathrm{b})}$ \\

\bottomrule
\end{tabular}

\vspace{3pt}
\begin{minipage}{\linewidth}
\footnotesize
\textsuperscript{$\dagger$} NT=4/8 \quad
\textsuperscript{$\ddagger$} NT=16/32 \quad
\textsuperscript{$\S$} FEDP-2K \\
\textsuperscript{*} Measured at FP8/BF8 precision;
2$\times$ at FP4,
$\frac{1}{2}\times$ at FP16/BF16,
and $\frac{1}{4}\times$ at TF32
\end{minipage}
\end{table}

\subsection{End-to-End Benchmarks}
\subsubsection{\textbf{TCU Extended SGEMM Kernel Roofline Analysis}}
\begin{figure}
    \centering
    \includegraphics[width=\linewidth]{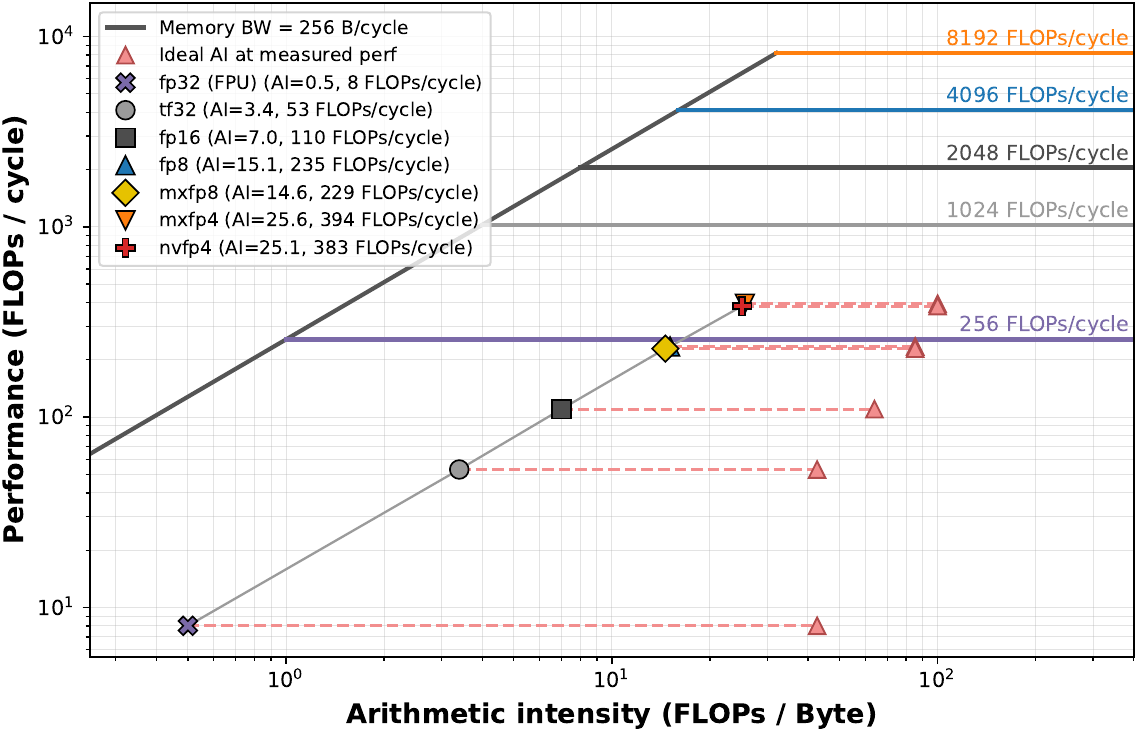}
    \caption{Ten-Four Enabled TCU-Extended SGEMM Roofline Analysis}
    \label{fig:roofline_256}
\end{figure}

We also perform roofline analysis on a $256{\times}256{\times}256$ SGEMM kernel across all Ten-Four supported formats, using a Vortex GPGPU configuration that is equivalent to a single NVIDIA SM (Fig.~\ref{fig:roofline_256}). The FPU-only FP32 baseline only achieves 8~FLOPs/cycle, whereas Ten-Four enabled TCUs reach up to 53, 110, 235, 394~FLOPs/cycle with TF32, FP16/BF16, FP8/BF8 and MXFP4 respectively. Throughput roughly doubles each time we halve operand width down to 8-bits but slows down to $\sim$1.7$\times$ at 4-bits due to block-scale-factor packing and metadata access overhead. Lower-precision operands also raise the kernel's achieved arithmetic intensity from 0.5~FLOPs/byte for the FPU baseline to 25.6 FLOPs/byte with MXFP4, a $\sim$51$\times$ increase that brings every TCU format to within 78--94\% of its ridge point. Larger tiles and improved operand reuse could further push these kernels past the ridge point and into the compute-bound region.

\subsubsection{\textbf{TCU Accelerated Application Kernels}} 
\begin{figure}
    \centering
    \includegraphics[width=\linewidth]{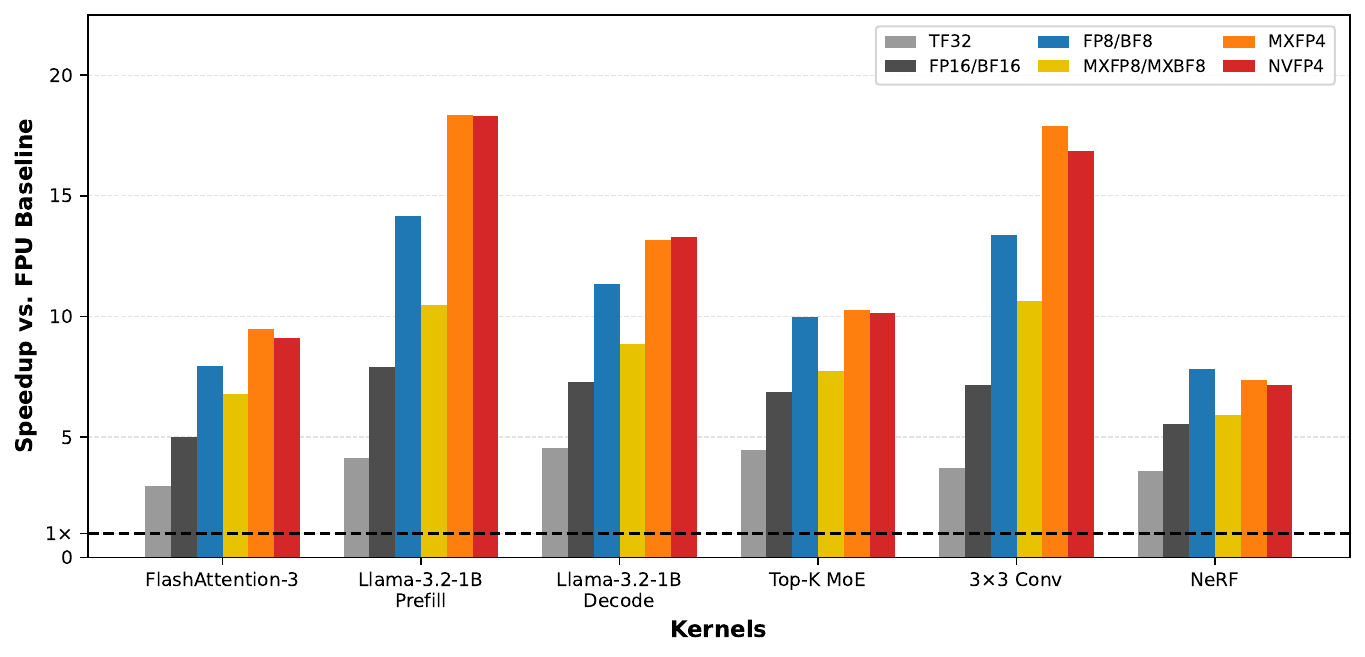}
    \caption{Ten-Four Enabled TCU-Extended End-to-End Application Kernel Speedup over FPU-Only Baseline}
    \label{fig:tfr_e2e}
\end{figure}

Finally, Table~\ref{tab:e2e_kernels} summarizes our evaluated end-to-end application kernels and Fig.~\ref{fig:tfr_e2e} shows per-format TCU speedups over an FPU-only baseline. We observe speedups across all kernels as we lower precision, from 3.0--4.5$\times$ with TF32 to 7.2--18.3$\times$ with the 4-bit formats. GEMM-dominated kernels benefit most, with Llama-3.2-1B prefill and $3{\times}3$ kernel convolution with im2col achieving 18.3$\times$ and 17.9$\times$ speedups with MXFP4. In contrast, kernels with substantial SIMT work (online softmax in FlashAttention-3, routing in MoE, and compositing in NeRF) naturally plateau at 9.5$\times$, 10.3$\times$, and 7.8$\times$ respectively.

\begin{table}
\centering
\small
\caption{Evaluated End-to-End Application Benchmarks}
\label{tab:e2e_kernels}

\setlength{\tabcolsep}{2.5pt}
\renewcommand{\arraystretch}{1.10}

\begin{tabular}{@{}p{1.95cm} p{3.95cm} p{2.35cm}@{}}
\toprule
\textbf{Kernel} &
\textbf{Description} &
\textbf{Parameters} \\
\midrule

FlashAttention-3 &
Fused $QK^\top$/$PV$ attention on TCU with online softmax,
on-device $P$ quantization; DMA producer warp &
2 heads, 768-token seq,
$d{=}128$, $B_M{=}48$ \\

\midrule

$3{\times}3$ Conv &
ResNet basic block with two convs:
im2col + TCU GEMM, fused bias/ReLU/residual &
$32{\times}32{\times}128$ NHWC,
$1024{\times}128{\times}1152$ GEMM\\

\midrule

NeRF &
Ray sampling and positional encoding, 10-layer TCU MLP,
SIMT RGB/density head and compositing &
32 rays ${\times}$ 32 samples,
width 256 \\

\midrule

MoE &
Top-$k$ gated FFN with SIMT routing, token permutation,
and grouped expert GEMMs with fused SwiGLU, and output combine &
512 tokens, $d{=}256$,
$d_{ff}{=}512$, 8 experts, top-2 \\

\midrule

Llama-3.2-1B\newline Prefill &
RMSNorm, QKV/O/FFN GEMMs on TCU,
RoPE, and FP32 GQA attn &
128 tokens \\

\midrule

Llama-3.2-1B\newline Decode &
Same transformer layer with batched autoregressive decode &
32 sequences,
512-token KV cache \\

\bottomrule
\end{tabular}
\end{table}

\section{Case Study of Custom Format Prototyping}
Beyond its FEDP microarchitecture, Ten-Four also serves as a framework for prototyping custom numerical formats and evaluating their associated circuitry's tradeoffs within a complete GPGPU Tensor Core datapath.

\subsection{LNSF4 Format}
As a demonstration, we propose LNSF4, a variation of NVIDIA's NVFP4 format~\cite{alvarez2025nvfp4, nvidia2024blackwell} that keeps the same FP4(E2M1) elements, 16-element block size, and FP32 tensor-wise scale, but replaces the FP8 (E4M3) scale-factor with an 8-bit fixed-point logarithmic number system (LNS8) encoding instead, following BOLD-Q \cite{11539596}. As with FP8 scale-factors, the MSB sign bit is redundant here since it only represents magnitude. The following 7 bits are interpreted as signed Q4.3 fixed-point, representing the value of $2^{E_{i}.E_{f}}$.

\subsection{Perplexity Analysis}
We first evaluate how well LNSF4 preserves model quality relative to NVFP4 by measuring perplexity on the WikiText-2~\cite{merity2016pointersentinelmixturemodels} and C4~\cite{JMLR:v21:20-074} benchmarks across multiple local LLMs from the Llama3~\cite{dubey2024llama3herdmodels} and Qwen3~\cite{yang2025qwen3technicalreport} families using the RaZeR \cite{chen2026razerpushinglimitsnvfp4} framework. We compare FP and LNS scales across three encoding variants: (1) Expanding the exponent/integer field to 5-bits by making use of the redundant sign bit (UE5M3 in FP, as adopted by NVIDIA's upcoming Rubin Architecture\cite{alvarez2026rubin, nvidia_ptx_isa_9_4}, and Q5.3 in LNS); (2) Expanding the mantissa/fraction field to 4-bits instead (UE4M4 and Q4.4); and (3) RaZeR's custom special-value element encodings (W4.3-A4.3 with SV$=\{\pm5\}$ and W3.3-A4.3 with SV$=\{\pm5,\pm8\}$) with the standard E4M3 or Q4.3 scale-factors. As shown in Table~\ref{tab:lns_g16_ppl}(a), the LNS variants closely track the FP ones across all 50 model/dataset/format combinations, with relative perplexity deviations only ranging from $-0.58\%$ to $+1.43\%$ ($0.28\%$ mean absolute difference). They even outperform FP in 28 of the 50 cases.

\begin{table*}[t]
\centering
\caption{Perplexity ($\downarrow$) and FPGA FEDP Synthesis Results for W4A4 Quantization with LNS vs FP 8-bit Scale-Factors} 
\label{tab:lns_g16_ppl}
\footnotesize
\setlength{\tabcolsep}{3.1pt}
\renewcommand{\arraystretch}{1.08}

\begin{tabular}{l
cc
cc
cc
cc
cc
c
ccc
c}
\toprule

&
\multicolumn{11}{c}{\textbf{\small (a) Perplexity Analysis}} &
\multicolumn{4}{c}{\textbf{\small (b) FPGA Synthesis}}
\\

\cmidrule(lr){2-12}
\cmidrule(lr){13-16}

\multirow{2}{*}{\textbf{Format}} &
\multicolumn{2}{c}{\textbf{Llama-3.1-8B}} &
\multicolumn{2}{c}{\textbf{Llama-3.2-3B}} &
\multicolumn{2}{c}{\textbf{Qwen3-4B}} &
\multicolumn{2}{c}{\textbf{Qwen3-8B}} &
\multicolumn{2}{c}{\textbf{Mistral-7B}} &
\multirow{2}{*}{\makecell{\textbf{LNS Wins}\\\textbf{on PPL}}} &
\multirow{2}{*}{\textbf{LUT}} &
\multirow{2}{*}{\textbf{FF}} &
\multirow{2}{*}{\makecell{\textbf{Dynamic}\\\textbf{Power (mW)}}} &
\multirow{2}{*}{\makecell{\textbf{LNS Wins}\\\textbf{on HW}}}
\\[-1pt]

\cmidrule(lr){2-3}
\cmidrule(lr){4-5}
\cmidrule(lr){6-7}
\cmidrule(lr){8-9}
\cmidrule(lr){10-11}

&
\textbf{Wiki} & \textbf{C4}
& \textbf{Wiki} & \textbf{C4}
& \textbf{Wiki} & \textbf{C4}
& \textbf{Wiki} & \textbf{C4}
& \textbf{Wiki} & \textbf{C4}
&
&
&
&
&
\\

\midrule

BF16
& 6.2403 & 8.9577
& 7.8166 & 10.4354
& 13.6625 & 16.6436
& 9.7265 & 13.3021
& 5.3182 & 7.8303
& --
& 4907 & 675 & 80
& --
\\

\midrule

FP 4.3
& \textbf{6.9343} & 9.9305
& 8.6059 & 11.6485
& \textbf{13.9723} & 17.2758
& \textbf{10.0502} & \textbf{13.7714}
& \textbf{5.5513} & \textbf{8.0909}
& \multirow{2}{*}{4/10}
& 2561 & 298 & 93
& \multirow{2}{*}{\textbf{3/3}}
\\

LNS 4.3
& 6.9358 & \textbf{9.9228}
& \textbf{8.5837} & \textbf{11.6111}
& 14.0909 & \textbf{17.2427}
& 10.1940 & 13.8050
& 5.5521 & 8.0924
&
& \textbf{2409} & \textbf{289} & \textbf{89}
&
\\

\midrule

FP 5.3
& 6.9361 & 9.9316
& 8.6099 & 11.6522
& \textbf{13.9654} & 17.2787
& \textbf{10.0664} & \textbf{13.7772}
& 5.5537 & \textbf{8.0943}
& \multirow{2}{*}{\textbf{6/10}}
& 2550 & 304 & 94
& \multirow{2}{*}{\textbf{3/3}}
\\

LNS 5.3
& \textbf{6.9344} & \textbf{9.9254}
& \textbf{8.5839} & \textbf{11.6338}
& 14.1176 & \textbf{17.2538}
& 10.1846 & 13.8050
& \textbf{5.5514} & 8.0977
&
& \textbf{2405} & \textbf{289} & \textbf{83}
&
\\

\midrule

FP 4.4
& 6.9270 & 9.9142
& \textbf{8.6013} & 11.6216
& 14.0983 & \textbf{17.1835}
& \textbf{10.2078} & \textbf{13.7896}
& \textbf{5.5423} & 8.0834
& \multirow{2}{*}{5/10}
& 2643 & 292 & 112
& \multirow{2}{*}{\textbf{3/3}}
\\

LNS 4.4
& \textbf{6.9175} & \textbf{9.8779}
& 8.6015 & \textbf{11.6126}
& \textbf{14.0172} & 17.2774
& 10.2715 & 13.8344
& 5.5459 & \textbf{8.0800}
&
& \textbf{2409} & \textbf{289} & \textbf{91}
&
\\

\midrule

RaZeR FP$_{\mathrm{W4.3-A4.3}}$
& 6.7774 & 9.7070
& 8.4225 & 11.3446
& \textbf{14.2906} & 17.3150
& \textbf{9.9514} & \textbf{13.6152}
& 5.4998 & 8.0267
& \multirow{2}{*}{\textbf{7/10}}
& 3045 & 314 & 132
& \multirow{2}{*}{\textbf{3/3}}
\\

RaZeR LNS$_{\mathrm{W4.3-A4.3}}$
& \textbf{6.7639} & \textbf{9.6827}
& \textbf{8.4020} & \textbf{11.3189}
& 14.2931 & \textbf{17.2470}
& 10.0029 & 13.6339
& \textbf{5.4975} & \textbf{8.0261}
&
& \textbf{2878} & \textbf{313} & \textbf{117}
&
\\

\midrule

RaZeR FP$_{\mathrm{W3.3-A4.3}}$
& 6.7608 & \textbf{9.6673}
& 8.4078 & 11.3119
& \textbf{14.1709} & 17.3814
& \textbf{9.8881} & \textbf{13.6878}
& 5.4974 & 8.0277
& \multirow{2}{*}{\textbf{6/10}}
& 3403 & 319 & 130
& \multirow{2}{*}{\textbf{3/3}}
\\

RaZeR LNS$_{\mathrm{W3.3-A4.3}}$
& \textbf{6.7542} & 9.6690
& \textbf{8.3881} & \textbf{11.3034}
& 14.2541 & \textbf{17.2855}
& 9.9453 & 13.7021
& \textbf{5.4909} & \textbf{8.0225}
&
& \textbf{3145} & \textbf{309} & \textbf{118}
&
\\

\bottomrule
\end{tabular}
\end{table*}

\subsection{LNSF4 Microarchitecture}

Having established LNSF4's numerical stability, we prototype it on Vortex using the RoadBlock framework \cite{rout2026roadblockrethinkinggputensor}, which provides the software quantization functions, host-kernel ABI, and MX metadata storage and tiling infrastructure required for end-to-end modeling of custom MX formats on GPGPU Tensor Cores. 

Both LNSF4 and NVFP4 share the exact same element processing datapath, comprising E2M1 to INT decoding, $4{\times}4$ element multipliers, and lane-grouped reduction per scale slot. The only difference arises in the block scale-factor compensation logic. Since NVFP4 requires a complete FP8 (E4M3) multiplication, it involves a $4{\times}4$ mantissa multiplier $(1.m_a)(1.m_b)$, a 5-bit 3-operand exponent adder, and a $12{\times}8$ multiplier for applying the 8-bit scale-factor product mantissa to the 12-bit lane-wise dot product. On the other hand, LNSF4 simply combines the two Q4.3 scales with a single 8-bit signed adder. The integer part of the result is fed into the element exponent adder, exactly as is done with standard E8M0 scale-factor MX formats, while the 3-bit fractional part of the result is used to index a tiny 8-entry antilog ROM storing $2^{k/8}$ in $[1.b]$ fixed point. The selected LUT entry is simply multiplied with the lane-grouped FP4 dot-product to produce the lane's final scaled mantissa. Additionally, the constant leading 1 allows us to avoid instantiating the leading row's AND-gate array in the already smaller $12{\times}(b{+}1)$ scale-application multiplier. 

\subsection{LNSF4 Synthesis and Optimization}
\begin{figure}
    \centering

    \subfloat[LUT Area Tradeoff\label{fig:lnsf4_allut_area}]{%
        \includegraphics[width=0.495\columnwidth]{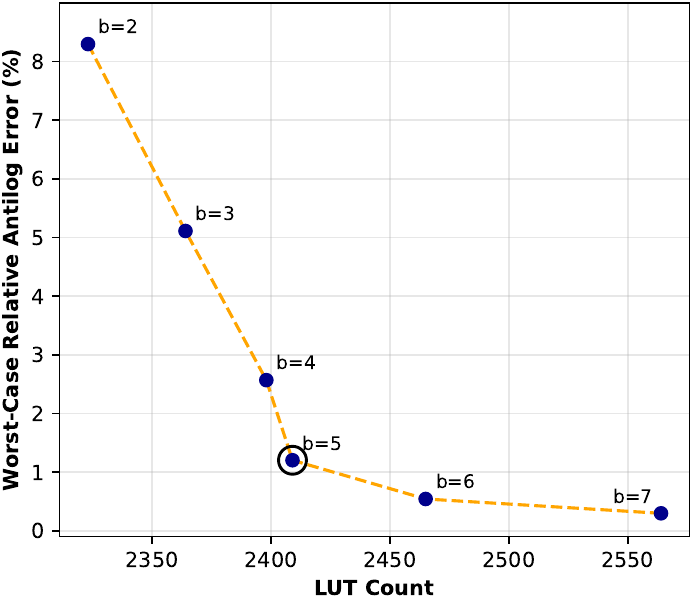}%
    }%
    \hfill
    \subfloat[Dynamic Power Tradeoff\label{fig:lnsf4_allut_power}]{%
        \includegraphics[width=0.495\columnwidth]{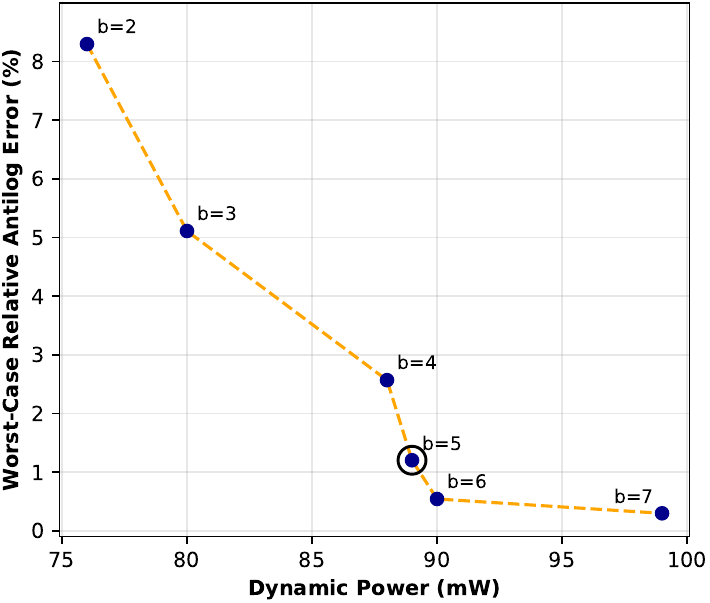}%
    }

    \caption{Worst-Case $\textbf{E}_{Frac}$ Antilog Error and Hardware Cost across LUT Precision Settings}
    \label{fig:lnsf4_allut_tradeoff}
\end{figure}

As shown in Table~\ref{tab:lns_g16_ppl}(b), LNSF4 consistently reduces hardware cost across all five FP/LNS variant pairs, achieving 5.5--8.9\% fewer LUTs, up to 4.9\% fewer FFs, and up to 18.8\% lower dynamic power. Furthermore, the fixed-point precision of the antilog LUT provides an additional opportunity to trade numerical accuracy for hardware cost, as explored in prior work on hardware-efficient antilogarithm conversion~\cite{4693995, nvidia2024logarithmic}. We sweep the number of fractional precision bits $b$ from 2 to 7 and evaluate the worst-case relative antilog error tradeoff against FPGA resource utilization and dynamic power consumption, as illustrated in Fig.~\ref{fig:lnsf4_allut_tradeoff}. Notably, increasing precision from 4 to 5 bits more than halves the approximation error with minimal additional hardware overhead, whereas further precision increases offer diminishing
accuracy improvements relative to their hardware cost. We therefore select $b{=}5$ as the accuracy--hardware Pareto-optimal point, achieving approximately 1\% worst-case relative antilog error with only 2409 LUTs, 289 FFs, and 89~mW dynamic power, and demonstrating Ten-Four's versatility and effectiveness as a platform for custom numerical format exploration.

\section{Conclusion}

In this paper, we introduced Ten-Four, a high-performance open-source Fused Dot Product Unit microarchitecture for developing a feature-rich mixed-precision Tensor Core Unit Extension to the RISC-V based Vortex GPGPU. By fusing integer and floating-point datapaths, supporting Microscaling with early addend accumulation, and exploiting dual-side sparsity through lane clock-gating, Ten-Four addresses the latency and resource inefficiencies of discrete arithmetic unit-based designs. Its configurable multiplier backends, pipeline depth, and routing optimizations further enable efficient mapping onto FPGA fabrics. On the AMD Alveo U55C FPGA, Ten-Four achieves 4-cycle latency at 300 MHz, delivering 130.368 GFLOPS peak throughput per Tensor Core while matching NVIDIA Tensor Core numerical accuracy. It achieves 2.7$\times$--7.9$\times$ speedup over discrete arithmetic library baselines at less than 60\% area cost, with 3.0$\times$--18.3$\times$ end-to-end kernel speedup over an FPU-only GPU implementation. Finally, our LNSF4 case study demonstrates Ten-Four's extensibility to custom numerical formats, achieving comparable perplexity to NVFP4 while reducing LUT utilization by 6.7\% and dynamic power by 11.1\%. Together, these results establish Ten-Four as a concrete solution for developing and evaluating next-generation mixed-precision GPGPU Tensor Cores.

\bibliographystyle{ACM-Reference-Format}
\bibliography{references}

\end{document}